\documentclass[sigconf,nonacm]{acmart}
\pdfoutput=1

\usepackage{booktabs}
\usepackage{graphicx}
\usepackage{balance}
\usepackage{enumitem}
\usepackage{xspace}
\usepackage[bf,BF,tight,small]{subfigure}
\usepackage[tableposition=top,font={scriptsize},figurename=Fig., skip=7pt]{caption}

\newcommand{\eg}{e.g.,\xspace}
\newcommand{\etc}{etc.\xspace}
\newcommand{\etal}{et al.\xspace}
\newcommand{\ie}{i.e.,\xspace}

\newcommand{\identifiers}{Publisher-specific\,IDs\xspace}
\newcommand{\publisherIDs}{\texttt{Publisher\,IDs}\xspace}
\newcommand{\containerIDs}{\texttt{Container\,IDs}\xspace}
\newcommand{\trackingIDs}{\texttt{Tracking\,IDs}\xspace}
\newcommand{\measurementIDs}{\texttt{Measurement\,IDs}\xspace}

\newcommand{\adstxt}{\texttt{ads.txt}\xspace}
\newcommand{\sellersjson}{\texttt{sellers.json}\xspace}
\newcommand{\covid}{COVID-19\xspace}

\newcommand{\trancoCrawl}{1MT crawl\xspace}
\newcommand{\point}[1]{\par\smallskip\noindent\textbf{#1}}

\makeatletter
\renewcommand\subsubsection{\@startsection
  {subsubsection}{0}{0mm}
  {0.5\baselineskip}
  {0\baselineskip}
  {\bf\normalsize\itshape}}
\makeatother

\begin{document}

\title[Who Funds Misinformation? A Systematic Analysis of the Ad-related Profit Routines of Fake News Sites]{Who Funds Misinformation? A Systematic Analysis of the Ad-related Profit Routines of Fake News Sites}

\subtitle{This work is published in WWW '23. Please cite DOI: 10.1145/3543507.3583443}

\author{Emmanouil Papadogiannakis}
\affiliation{
	\institution{FORTH \& University of Crete}
	\country{Greece}
}
\author{Panagiotis Papadopoulos}
\affiliation{
	\institution{FORTH}
	\country{Greece}
}
\author{Evangelos P. Markatos}
\affiliation{
	\institution{FORTH \& University of Crete}
	\country{Greece}
}
\author{Nicolas Kourtellis}
\affiliation{
	\institution{Telefonica Research}
	\country{Spain}
}

\renewcommand{\shortauthors}{Emmanouil Papadogiannakis, Panagiotis Papadopoulos, Evangelos P. Markatos, \& Nicolas Kourtellis}

\begin{CCSXML}
<ccs2012>
   <concept>
       <concept_id>10002951.10003260.10003272</concept_id>
       <concept_desc>Information systems~Online advertising</concept_desc>
       <concept_significance>500</concept_significance>
       </concept>
   <concept>
       <concept_id>10002951.10003260.10003277.10003281</concept_id>
       <concept_desc>Information systems~Traffic analysis</concept_desc>
       <concept_significance>500</concept_significance>
       </concept>
   <concept>
       <concept_id>10002951.10003260.10003277.10003279</concept_id>
       <concept_desc>Information systems~Data extraction and integration</concept_desc>
       <concept_significance>500</concept_significance>
       </concept>
 </ccs2012>
\end{CCSXML}

\ccsdesc[500]{Information systems~Online advertising}
\ccsdesc[500]{Information systems~Traffic analysis}
\ccsdesc[500]{Information systems~Data extraction and integration}

\keywords{Fake News, Online Advertising, Web Monetization}

\begin{abstract}
Fake news is an age-old phenomenon, widely assumed to be associated with political propaganda published to sway public opinion.
Yet, with the growth of social media, it has become a lucrative business for Web publishers.
Despite many studies performed and countermeasures proposed, unreliable news sites have increased in the last years their share of engagement among the top performing news sources.
Stifling fake news impact depends on our efforts in limiting the (economic) incentives of fake news producers.

In this paper, we aim at enhancing the transparency around these exact incentives, and explore:
Who supports the existence of fake news websites via paid ads, either as an advertiser or an ad seller?
Who owns these websites and what other Web business are they into?
We are the first to systematize the auditing process of fake news revenue flows.
We identify the companies that advertise in fake news websites and the intermediary companies responsible for facilitating those ad revenues.
We study more than 2,400 popular news websites and show that well-known ad networks, such as Google and IndexExchange, have a \emph{direct advertising} relation with more than 40\% of fake news websites.
Using a graph clustering approach on 114.5K sites, we show that entities who own fake news sites, also operate other types of websites pointing to the fact that owning a fake news website is part of a broader business operation. 
\end{abstract}

\maketitle

\section{Introduction}
\label{sec:introduction}

Misinformation, or formally, spreading ``incorrect or misleading information''~\cite{lazer2018science}, is not a new phenomenon, but the tools people use to spread misinformation have dramatically improved with the Internet and social media.
Fake news and misinformation, not only pose serious threats to the integrity of journalism, but have also created societal turmoils
in the economy~\cite{ethereumDeath}, the political world~\cite{10.1177/0894439317734157,teensInBalkans} and even in human life~\cite{whoStudy}.
Unlike the yellow newspapers of the past that have been capitalizing on fake news for decades, social media and search engines pose an additional threat to truth: the more luring the content of a website is, the more it is promoted by the algorithms underpinning these platforms. 
BBC interviewed 50 experts about the ``grand challenges of the 21st century'' and many of them named propaganda and fake news~\cite{bbc} as a key challenge. 

Considering its significant impact, tech firms, researchers, governments and stakeholders have explored various methods to identify and curtail the spread of fake news.
Google and other tech companies, signed up to a voluntary EU code of conduct which required them to ``improve the scrutiny of ad placements to reduce revenues of the purveyors of disinformation''~\cite{gni}.
Also, there is an abundance of academic works aiming at analyzing~\cite{agarwal2020stop,lazer2018science,10.1145/3309699,chalkiadakis2021rise,potthast2017stylometric} or detecting~\cite{perez2017automatic,zhang2020overview,zhou2019fake,shu2017fake} fake news sources on the Web.

Despite these important actions, unreliable news sites significantly increased ($2.1\times$) their share of engagement among top performing news sources in the past year alone~\cite{traction}.
The success of curbing fake news primarily depends on the efforts to reduce or even eliminate the incentives of fake news producers.
But, admittedly, there is little we know about the incentives and funding of fake news on the Web. 
Aside from various political gains that may motivate the spread of doctored narratives~\cite{agarwal2020stop,buzzfeed}, disseminating fake news has been a lucrative Web business~\cite{8123490}.
The ad industry provides wide avenues for high revenues: for every \$2.16 spent on news websites in USA, \$1 is spent on misinformation~\cite{revenues1,revenues2}.
In fact, ad-tech agencies intensely track~\cite{englehardt2016online, papadogiannakis2021user} and programmatically bid~\cite{10.1145/3131365.3131397,olejnik2013selling,10.1145/3355369.3355582} for ad spaces (of lower cost) that reside in websites of questionable content.
Thus, ad budgets move from high quality news websites to low-cost, controversial ones~\cite{adfoo}, with various examples of ads from prestigious companies (\eg Microsoft, Citigroup, IBM) and small business owners being placed on (and thus unwittingly funding) websites that promote fake or even illegal (\eg Jihadi~\cite{jihad} and neo-nazi~\cite{nazi} related) content.

In this study, we shed light on the revenue flows of fake news websites by investigating who supports and maintains their existence.
We do not examine what misinformation is, rather, we investigate who provides revenue to fake news websites.
We systematize the auditing process of digital advertising in those websites by developing a  methodology, which enables us to identify 
(i) the intermediary companies that sell the ad space of fake news websites to the ad ecosystem, 
(ii) the advertisers who buy the ad space on such sites, and
(iii) the type of ads they place.

The contributions of this paper are summarized as follows:
\begin{enumerate}
    \item We develop a novel ad detection methodology which enables us to identify the advertisers that collaborate with fake news websites. We find that about 70\% of the fake news websites advertise ``Business'' products and services, and close to 40\% display ``Entertainment'' advertisements.
    
    \item We study who provides the ad revenues of fake news websites and show that the most well-known legitimate advertising networks (such as google.com, indexexchange.com, and appnexus.com) have a \emph{direct} advertising relation with more than 40\% of the fake news websites in our dataset, and have a \emph{reseller} relation with more than 60\% of those sites.

    \item We show that owners of fake news websites own other types of websites as well, including ``Entertainment'', ``Business'', and ``Politics''.
    This implies that the operation of an average fake news website is not an isolated or outlying event, but instead is probably part of a wider business function. 
    
    \item We make our lists of fake and real news websites, ad creatives collected on top 100 websites, fake news clusters, and code of crawler and ad detection method publicly available~\cite{openSourceData,openSourceCode}.
\end{enumerate}
\section{Related Work}
\label{sec:related}

\point{Fake News:}
There has been a lot of effort to create datasets that enable future research on misinformation~\cite{murayama2021dataset}.
Most recently, in~\cite{kim2021fibvid}, authors produced a dataset regarding fake news information related to the \covid pandemic, while in~\cite{patwa2021fighting}, authors manually annotated news articles and social media posts of real or fake \covid stories.
In~\cite{golbeck2018fake}, authors collected and evaluated news articles regarding American Politics, resulting in a dataset of fake news and satirical articles, along with a factual article that disproves them.
In~\cite{zhou2020recovery}, authors analyzed over 2K news articles and 140K tweets on the \covid pandemic, formed lists of reliable or unreliable news publishers, and explored spread of \covid articles on Twitter.

Similar to our work, in~\cite{bozarth2021market} authors explored the advertising market of traditional, fake news and low-quality news websites.
Using a manually curated list of popular ad servers, they found that fake publishers rely on credible ad servers to display ads and monetize their traffic.
In~\cite{hounsel2020identifying}, authors utilized non-perceptual features (\eg domain name, DNS config) to train a multi-class model that detects disinformation websites in the wild.
In~\cite{han2021infrastructure}, Han~\etal studied how Web infrastructure supports misinformation and hate speech websites.
They found that fake news websites disproportionately rely on hosting providers (\eg Cloudflare), and that they mainly rely on Revcontent and Google to generate revenue. 

Bakir~\etal~\cite{bakir2018fake} discussed how the lack of understanding and control advertisers have regarding where their ads appear, enables fake news websites to generate revenue.
They explained how fake news websites can proliferate by moving to a new ad network once blocked in another.
Zeng~\etal~\cite{zeng2020bad} studied problematic ads (\eg clickbait, scams) and their prevalence across news websites, as well as the ad platforms that serve them.
Similar to our work, they discover that fake news websites work with the same ad platforms as real news websites and that similar ads are served in both categories.
However, contrary to our work, authors did not study the revenue flow associated with ads and only focus on the ad content.

Finally, the Global Disinformation Index often conducts studies to assess the ad companies that inadvertently facilitate misinformation websites~\cite{gdi1, gdi2}.
The Check My Ads Institute reviews the adtech industry and attempts to disrupt the revenue flows of disinformation and hate speech outlets~\cite{checkMyads}.
Similarly, the Sleeping Giants activism movement creates awareness regarding how ads are distributed across the ecosystem and has managed to reduce the ad revenue of fake news websites~\cite{li2021beyond,braun2019activism}.

\point{Website Administration:}
Academic research has focused on identifying the legal entities that control and operate websites.
The methodology followed in this work is closely related to the one presented in~\cite{coOwnershipGraphs}.
Specifically, authors proposed a graph-based model of website administration using ad network and tracking services relationships.
Through a large-scale analysis on the monetization models of ad networks and Web publishers, they detected patterns of preferential administration of websites.
In our work, we make use of the proposed Metagraph to detect websites operated or even owned by the same entity.
We refrain from analyzing the behavior of intermediary publishing partners since they do not provide any additional information to this work.
In~\cite{simeonovski2017controls}, authors studied security threats and the involved entities by making use of HTTPS certificates to extract organization names.
In~\cite{cangialosi2016measurement} the authors utilized email addresses found in WHOIS records to extract groups of domains owned by the same entity.

\point{Ad Detection:}
The study, detection and exclusion of advertisements in websites has been the focus of research work for a long period of time.
In~\cite{li2012knowing}, the authors proposed \emph{MadTracer}, a system that detects malicious advertisements in websites based on the redirection chains among publishers and ad networks.
In~\cite{bozarth2021market}, similar to our methodology, authors crawl websites and extract URLs embedded in the webpage and in iFrames.
Using EasyList, they form a list of popular ad servers, against which URLs are matched.
Contrary to our work, they do not examine network traffic and delivered content for ad detection.
In~\cite{iqbal2020adgraph}, authors presented \emph{AdGraph}, a graph-based system that detects ads and other tracking resources in websites.
AdGraph provides a graph representation of the website rendering, network traffic and Javascript execution.
In~\cite{sjosten2020filter}, the authors presented \emph{PageGraph}, a similar but more robust graph representation system.
In~\cite{siby2021webgraph} authors proposed \emph{WebGraph} which builds a graph representation of the webpage but focuses on the actions of ads instead of their content.
Contrary to these techniques, we do not focus on the rendering of websites or the execution of code, nor do we use a trained model.
Our methodology combines external block lists with network traffic monitoring, making it more agile to adapt, as it does not require (re)training models.
\section{Overview of Methodology}
\label{sec:methodology}

\begin{figure}[t]
    \centering
    \includegraphics[width=.97\columnwidth]{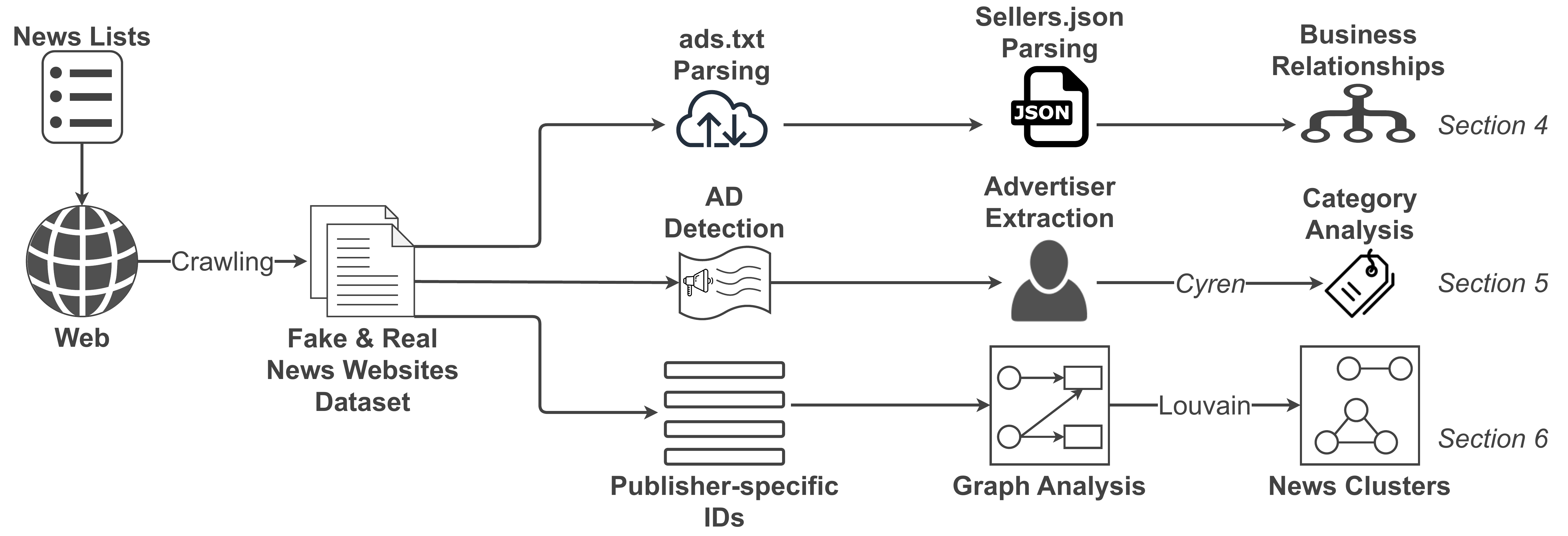}
    \caption{Overall methodology of the present study.}\vspace{-0.2cm}
    \Description{Abstract diagram of methodology followed in this work, where each horizontal component illustrates the steps followed in each main section of this work.}
    \label{fig:overallMethodology}
\end{figure}

In this section, we outline the methodological steps we follow to investigate fake news websites and the entities that support them.
As illustrated in Figure~\ref{fig:overallMethodology}, we first construct a list of fake and real news websites and crawl them to collect ad-related data on each.

\begin{figure*}[t]
   \begin{minipage}[t]{0.31\textwidth}
        \centering
        \includegraphics[width=\columnwidth]{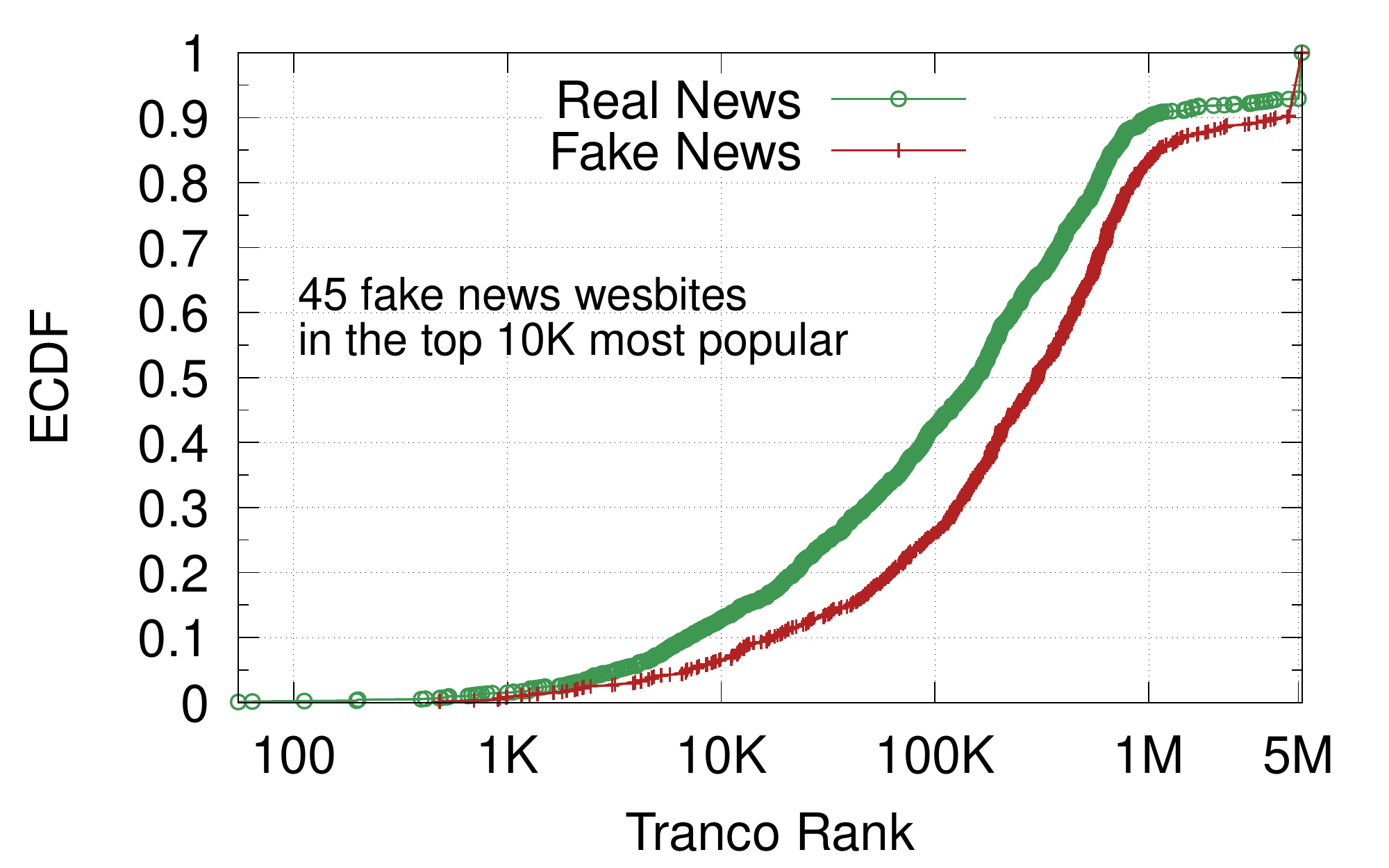}\vspace{-0.2cm}
        \caption{Distribution of rank of news websites based on Tranco. There are both fake and real news websites with very high popularity. In addition, we see that both lists contain websites of comparable popularity.}\vspace{-0.3cm}
        \Description{Cumulative Distribution Function of the popularity of fake news and real news websites, where we observe that the two sets have a similar distribution.}
        \label{fig:ranks}
   \end{minipage}
    \hfill
    \begin{minipage}[t]{0.31\textwidth}
        \centering
        \includegraphics[width=\columnwidth]{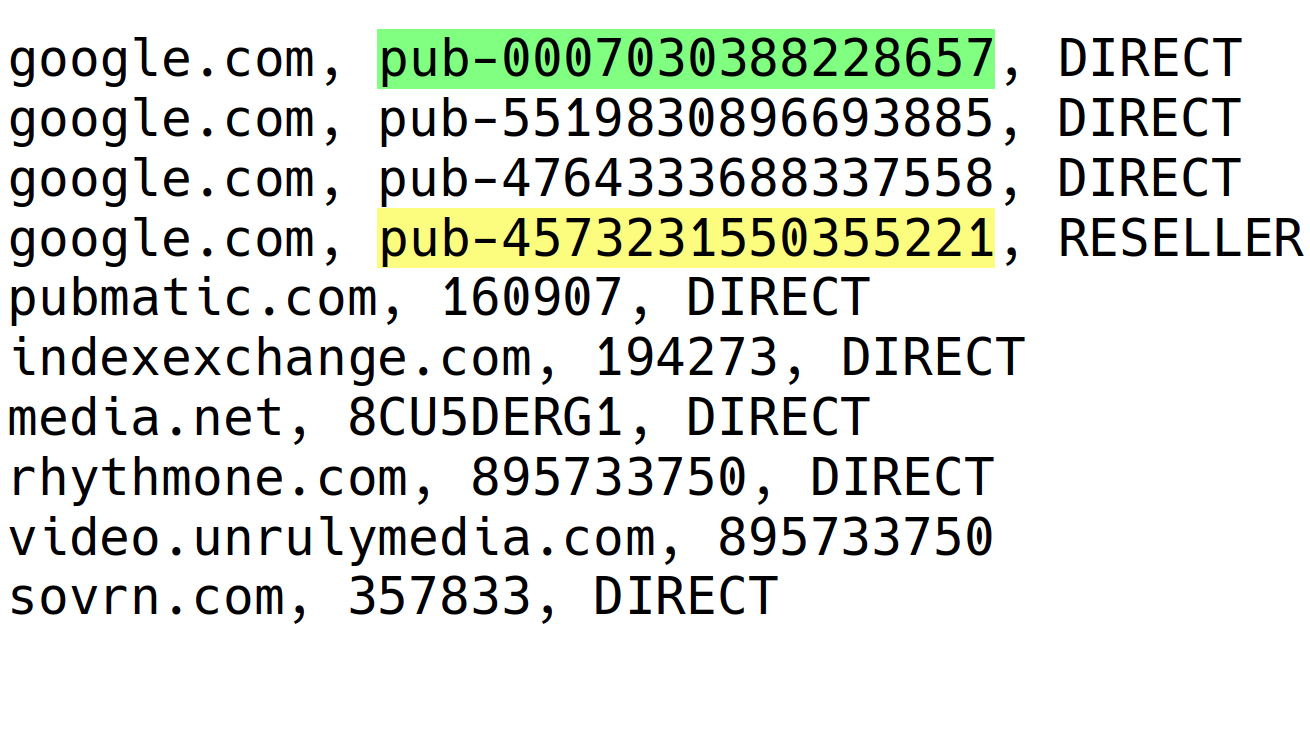}\vspace{-0.2cm}
        \caption{Snippet of the \adstxt file served by \emph{ainarsgames.com} on October 2022. Each entry is a unique business relationship. Identifiers can be matched against sellers.json files (see Fig~\ref{fig:sellersJson}).}
        \Description{Text that represents a snippet of an ads.txt file. There are multiple lines of text and each line is a distinct ads.txt entry. Each entry consists of comma-separated fields. Two publisher IDs, issued by Google are highlighted.}
        \label{fig:adsTxt}
    \end{minipage}
    \hfill
    \begin{minipage}[t]{0.31\textwidth}
        \centering
        \includegraphics[width=\columnwidth]{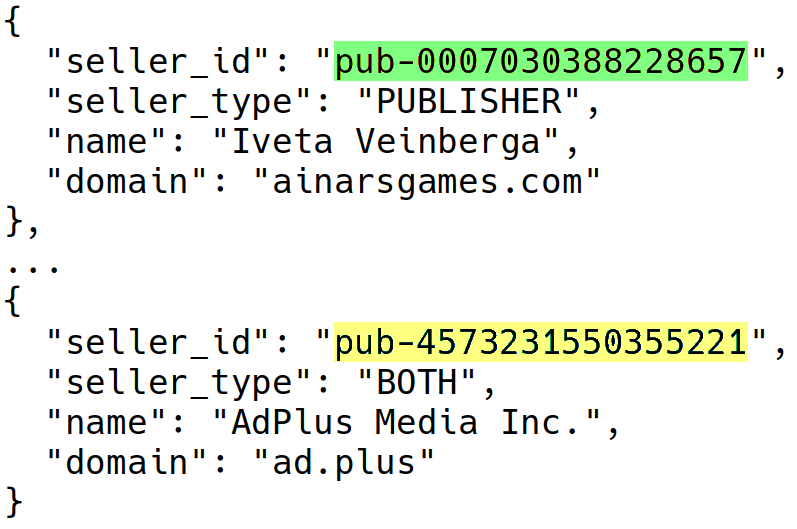}\vspace{-0.2cm}
        \caption{Snippet of the \sellersjson served by Google on October 2022. Each entry represents an entity with which Amazon has a business relationship.}
        \Description{Text that follows the JSON format and represents a snippet of a sellers.json file. There are two JSON objects that represent two different business relationships. Each object contains the "seller ID", the "seller type", the "name" and the "domain" keys. The values of the "seller ID" fields are highlighted and match those of the ads.txt figure.}
        \label{fig:sellersJson}
    \end{minipage}
    \vspace{-0.2cm}
\end{figure*}

\subsection{Fake and real news website lists}
\label{sec:lists}
We utilize publicly available datasets to create a corpus of fake and real news websites, and ensure the reproducibility of our work.
\begin{enumerate}[itemsep=0pt,topsep=0pt,leftmargin=0.5cm]
\item MediaBias/FactCheck (MBFC)~\cite{mbfc}: MBFC is an independent organization that aims to detect bias of media and other information sources by following a very strict manual methodology~\cite{mbfcMethodology}, that makes use of a combination of objective measures\footnote{Already used in numerous past studies~\cite{gruppi2021nela,chen2020proactive,ghanem2021fakeflow,chowdhury2020joint,zhou2020recovery}}. 
We download the list on 15 Nov 2021 and extract the websites that have been labeled as ``Questionable Source'' or ``Conspiracy-Pseudoscience'' and have ``Low'' or ``Very Low'' factual reporting and their credibility has been described as ``Low''.
These websites manifest extreme bias, obvious propaganda, lack of proper sourcing to credible information, complete lack of transparency, and focus on spreading fake news.
From the original list of 3,915 websites, we conclude with 816 fake news websites that meet the above criteria.

\item Columbia Journalism Review (CJR)~\cite{cjr}: CJR is a journal for professionals of various disciplines. Their list was created by merging some of the most common fake news lists (\eg OpenSources~\cite{opensources}, Politifact~\cite{politifactAlmanac} and Snopes~\cite{snopes}).
CJR curated the list to remove high-partisan websites that do not serve fake news content.
From the resulting list, we consider websites that have been labeled as ``fake news'', ``conspiracy'' or ``extremely biased websites'', and end up with a list of 350 websites\footnote{At the moment of this writing, it is accessible through an archive site~\cite{cjr}.}.

\item Golbeck \etal~\cite{golbeck2018fake}: In this dataset, authors focus on fake news and satirical articles related to USA politics, posted after January 2016.
They follow a manual investigation process, where each article is evaluated by two researchers.
Additionally, for fake news articles, they provide a link to a well-researched, factual article that rebutted the fake news story.
From this list, we select 55 websites that have been found to have published at least three such articles by both evaluators.
This threshold has been determined based on empirical analysis.

\item Zhou~\etal~\cite{zhou2020recovery}: Authors created a dataset of 2,029 news articles and 140,820 tweets related to \covid.
Regarding the news articles, they extracted knowledge using \emph{NewsGuard}~\cite{newsguard} and MBFC.
Similar to the above, we extract 31 websites that have published at least three fake stories.
\end{enumerate}

\noindent Real News: Additionally, we form a list of credible news websites that serve factual content, cite credible sources and usually cover both sides of reported stories.
We focus on websites that have been evaluated by MBFC and have been found to have minimal or no bias.
Specifically, we extract websites that have been labeled as ``Pro-Science'', ``Least-Biased'', ``Left-Center'' or ``Right-Center'' and have ``High'' or ``Very High'' factual reporting and ``High'' credibility.
We do not make any assumptions about the spread of misinformation across the political spectrum, however, MBFC uses the labels ``Left-Center'' and ``Right-Center'' for websites which are less biased and generally trustworthy.
This, in conjunction with the fact that we also require that they have been labeled with high factual reporting, ensures that such websites are credible.
This approach results in a list of 1,368 credible websites, which we refer to as \emph{real news}.

\begin{table}[t] 
    \centering
    \scriptsize
    \begin{tabular}{llr}
    \toprule
        \textbf{Source} & \textbf{Description} & \textbf{\# Websites} \\
    \midrule
        1. Media Bias/Fact Check~\cite{mbfc} & Questionable Sources & 816 \\
        2. CJR~\cite{cjr} & Fake News \& Biased & 350 \\
        3. Golbeck \etal~\cite{golbeck2018fake} & Fake \& Satire Articles & 55 \\
        4. Zhou \etal~\cite{zhou2020recovery} & News articles & 31 \\
        5. Media Bias/Fact Check~\cite{mbfc} & High Credibility & 1,368 \\
    \midrule
        \textbf{Total unique fake news websites}  &  &\textbf{1,044} \\
        \textbf{Total unique real news websites}  &  &\textbf{1,368} \\
    \bottomrule
    \end{tabular}
    \caption{Sources of fake and real news sites and unique total used.}\vspace{-0.3cm}
    \label{tab:fakenewsLists}
\end{table}

\point{Fake \& Real News Lists:}
By combining these sources, we construct a list of 1,044 unique fake news and 1,368 unique real news websites.
Please note that there is an overlap across fake news lists and there are websites which can be found in multiple lists.
For instance, the website \emph{infowars.com} has been labeled as a misinformation source in all 4 sources.
Table~\ref{tab:fakenewsLists} summarizes the aforementioned sources of fake and real news websites.
Our lists are publicly available~\cite{openSourceData}. 
To understand the popularity of the sites in our dataset, in Figure~\ref{fig:ranks} we plot their ranking based on the Tranco list~\cite{tranco}, from 18.10.2021 to 16.11.2021\footnote{https://tranco-list.eu/list/YKQG/full}.
We see that 45 fake news websites are among the top 10K most popular sites and such rankings usually translate in a wide audience with millions of visitors per month.
We observe that websites in the two lists have very similar rankings, suggesting that they attract a  similar number of visitors and therefore are directly comparable.

\subsection{Website crawling}
\label{sec:crawler}
We develop a puppeteer-based crawler that stores (i) the HTML content of the visited website, (ii) a cookie-jar for both first-party and third-party cookies, (iii) the \texttt{ads.txt} file (if present), (iv) a screenshot of the landing page, and (v) the HTTP(S) network traffic.
We also implement the ad-detection mechanism, described later in Section~\ref{sec:adDetection}.
The implementations of both the crawler and the ad detection methodology are publicly available~\cite{openSourceCode}.
Using this crawler, we visit the landing page of real and fake news websites on 13 Dec 2021.
The crawler was located in an EU-based institution and collected about 31GB of data.
The timeout for loading each website was set up to 60 seconds.
Ethical aspects of our study are discussed in Appendix~\ref{sec:ethics}.
\section{News Website Financing}
\label{sec:facilitators}

\subsection{Who sells ad space on fake news sites?}
\label{sec:adsTxt}

First, we study the entities selling ad space to understand who facilitates the monetization of news websites.
To achieve this, we utilize \adstxt files served by websites.
An {\bf \adstxt} file~\cite{adsTxtStandard} is a simple text file located at the root of a website that explicitly states which auctioneers are authorized to sell the impression inventory of this website.
In order for the entire ad ecosystem to work as expected, Supply-Side Platforms (SSPs) should ignore inventory which they are not authorized to sell, while Demand-Side Platforms (DSPs) should not buy inventory from unauthorized sellers.
As shown in Figure~\ref{fig:adsTxt}, each record in \adstxt is an entry with comma-separated fields and it authorizes a specific SSP to sell impressions for this website.
These fields are:
(i) the domain of the SSP,
(ii) an identifier which uniquely identifies the account of the publisher within the service
(iii) the relationship for this account (can be either DIRECT \ie the publisher is the owner of the specified account or RESELLER \ie a third party has been assigned by the website owner to manage the specified account), and
(iv) optionally, an identifier that maps to the company listed in (i) and uniquely identifies it within a certification authority.
Every such entry defines a business relationship between the owner of the website and the seller~\cite{bashir2019longitudinal}.

We parse and analyze the content of these files, fetched by the crawler described in Section~\ref{sec:crawler}.
In total, we find 198 fake news websites and 627 real news websites serving a valid \adstxt file that follows the specification~\cite{adsTxtStandard}.
According to specification~\cite{adsTxtStandard}, relationships of \texttt{DIRECT} type indicate that the publisher (\ie the owner of the content) directly controls the specific account in the respective service.
Consequently, these relationships are of special interest, since they disclose a direct business contract between the publisher and the ad network.
An analysis of the \texttt{RESELLER} relationships is presented in Appendix~\ref{sec:adNetworks}. 

\begin{figure*}[ht]
    \centering
    \hfill
    \begin{minipage}[t]{0.323\textwidth}
        \centering
        \includegraphics[width=\columnwidth]{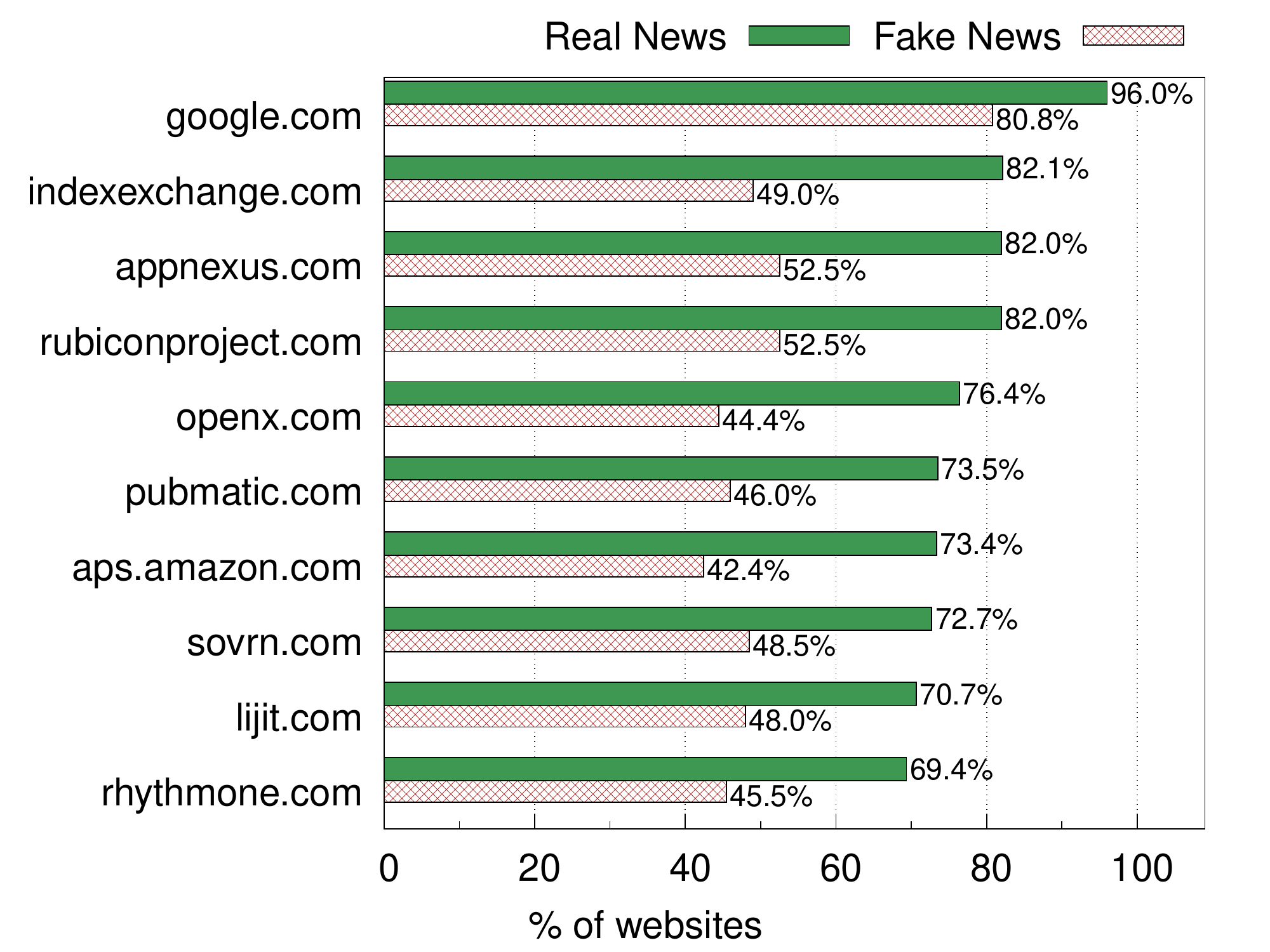}\vspace{-0.2cm}
        \caption{Most popular authorized digital sellers with \texttt{DIRECT} relationship in \adstxt files. Identifiers of such entries indicate that the publisher is the direct owner of the account. We observe that the majority of news websites have business relationship with Google.}\vspace{-0.3cm}
        \Description{Histogram of most popular ad networks that have a business relationships with news websites, as reported in ads.txt files. Each bar represents the percentage of news websites that discloses their relationship with the ad network in their ads.txt files. The ad networks are google.com, indexexchange.com, appnexus.com, rubiconproject.com, openx.com, pubmatic.com, aps.amazon.com, sovrn.com, lijit.com and rhythmone.com}
        \label{fig:adsTxtDirect}
    \end{minipage}
    \hfill
    \begin{minipage}[t]{0.323\textwidth}
        \centering
        \includegraphics[width=\columnwidth]{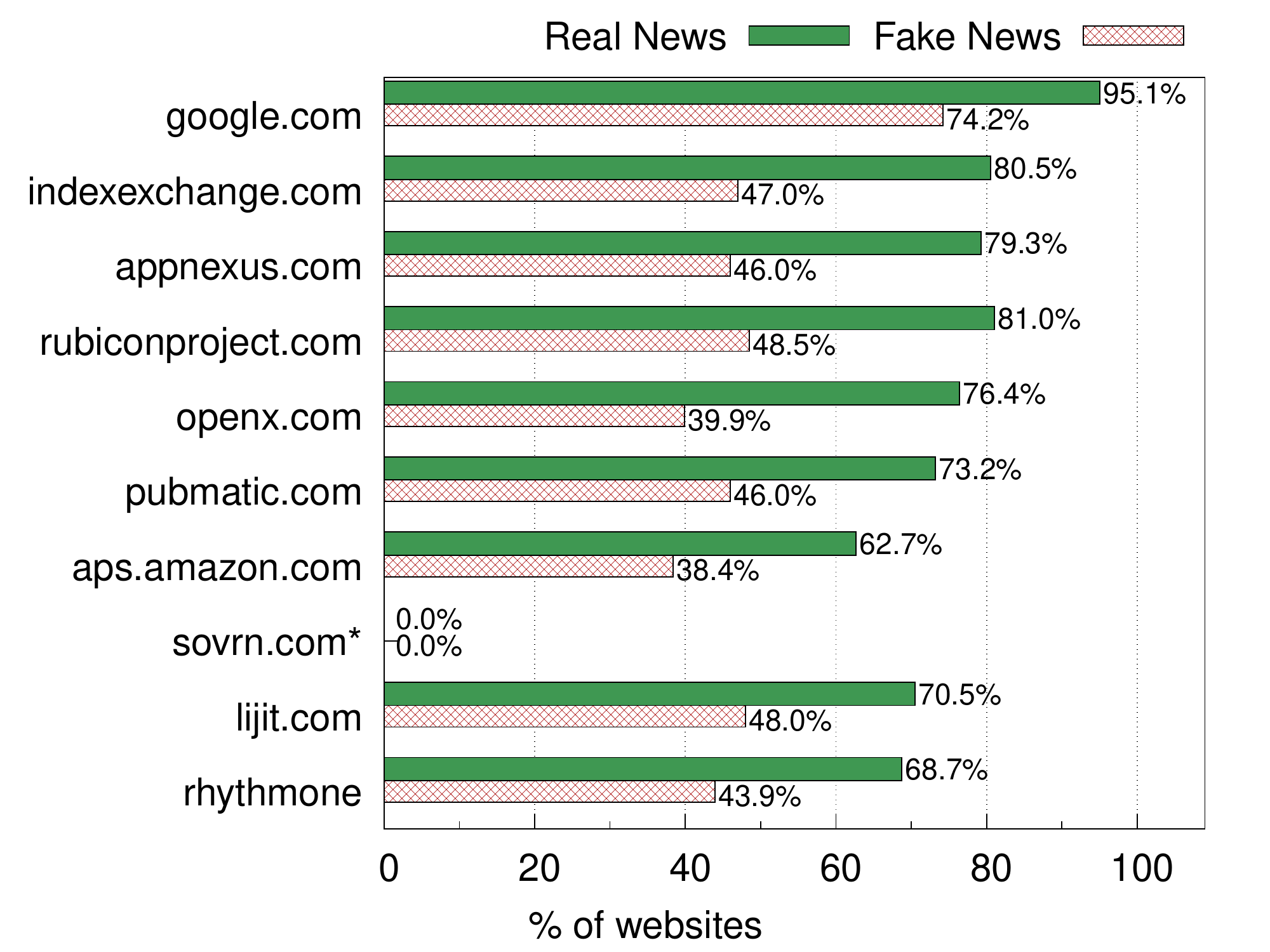}\vspace{-0.2cm}
        \caption{Direct business relationships between news websites and ad networks verified by entries in \adstxt and \sellersjson files. \emph{sovrn.com} is excluded since its \sellersjson file could not be retrieved.}\vspace{-0.3cm}
        \Description{Histogram of most popular ad networks that have business relationships with news websites, as reported in sellers.json files. Each bar represents the percentage of news websites that the ad network works with and discloses through their sellers.json files. The ad networks are google.com, indexexchange.com, appnexus.com, rubiconproject.com, openx.com, pubmatic.com, aps.amazon.com, sovrn.com, lijit.com and rhythmone.com}
        \label{fig:sellersJsonDirect}
    \end{minipage}
    \hfill
    \begin{minipage}[t]{0.323\textwidth}
        \centering
        \includegraphics[width=\columnwidth]{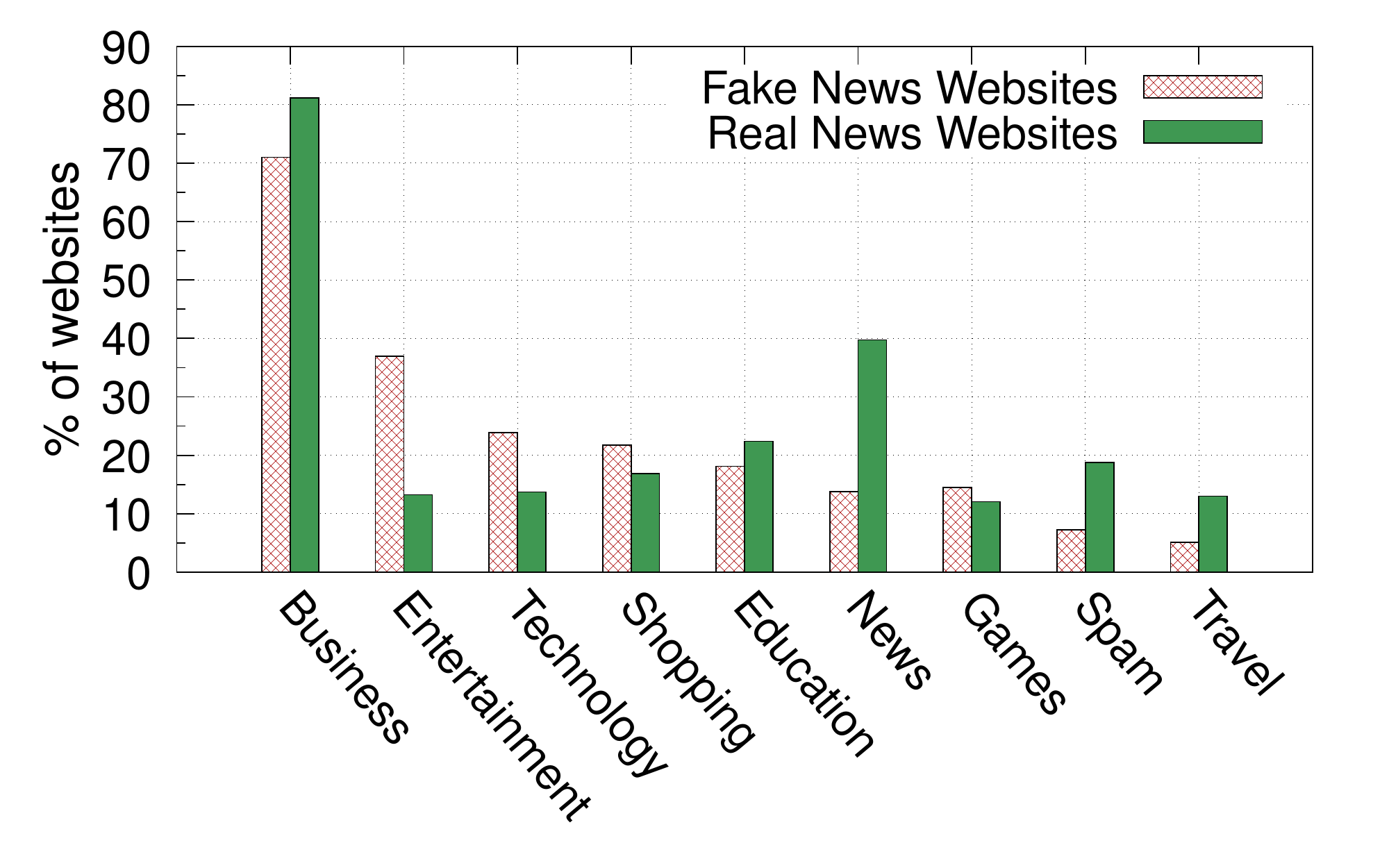}\vspace{-0.2cm}
        \caption{Distribution of categories of advertisers appearing in fake and real news websites. For both types the majority of advertisers provide business-related information.}
        \Description{Histogram of types of advertisers identified in news websites. Each bar represents the percentage of news websites that contain an advertisement of the respective type. There are bars for fake and real news websites. The types of advertisements are Business, Entertainment, Technology, Shopping, Education, News, Spam and Travel.}
        \label{fig:advertisersCategories}
    \end{minipage}
    \vspace{-0.2cm}
\end{figure*}

For each ad network, we measure the portion of websites that provide an \adstxt file and have a business relationship with it.
We find that, on average, fake news websites in our dataset form direct business relationships with 27 ad systems, while surprisingly, real news websites do so with 41 systems.
In Figure~\ref{fig:adsTxtDirect}, we illustrate the top 10 most popular digital sellers of ads for the \texttt{DIRECT} relationships that appear in both real and fake news websites (\ie intersection).
As the figure suggests, a large portion of real news websites tends to form \texttt{DIRECT} business relationships with well-known ad networks (\eg 96\% of real news with google.com, and 82.1\% with \emph{indexexchange.com}).
Even though ad platforms are found in these files, they might not end up serving any ads due to the nature of programmatic advertising.
However, there is still a business relationship between the website and the ad network.

What is more interesting, however, is that a lot of fake news websites also have direct business relationships with these ad networks. 
Indeed, 80.8\% of fake news websites have a direct business relationship with \emph{google.com}, 49\% of fake news websites with \emph{indexexchange.com}, and 52.5\% with \emph{appnexus.com}.
By independently examining the top ad systems for fake and real news websites, we find that \emph{revcontent.com} is the only ad system that is popular (\ie ranked 5\textsuperscript{th}) among the ad networks integrated with fake news websites, but ranked very low (\ie 51\textsuperscript{st}) among the ad networks of real news websites, which suggests that this network is preferred by fake news websites.
Contrary, we find \emph{yahoo.com} being preferred by real news websites: 68\% of them form a business relationship with \emph{yahoo.com}, while only 30\% of fake news websites do so.

We observe that only a portion of fake news websites in our list provide an \adstxt file.
We recrawl our list of fake news websites on January 31, 2023 and find that 262 websites now serve \adstxt files (up from 198).
We find similar results, with the top ad-networks being almost identical.
83.9\% of fake news websites have a \texttt{DIRECT} relationship with Google, 47.32\% with IndexExchange, \etc
Studying the third parties that fake news websites interact with, shows that for the 198 websites serving \adstxt files, 94.95\% of them interact with Google-owned tracking or ad-serving domains.
We classify domains as trackers based on the list provided by Disconnect~\cite{disconnectList}.
Looking into all the crawled fake news websites, regardless if they serve \adstxt files, we find 84.08\% of them interacting with Google.
The above support our findings that
(i) the fake news websites with \adstxt files we studied are representative of the ad-ecosystem;
(ii) popular ad systems provide ad revenue to fake news websites.

\point{Finding:} Although the percentages vary from one ad network to the next, Figure~\ref{fig:adsTxtDirect} suggests that, on average,  popular ad networks have \texttt{DIRECT} business relationship with about half of the fake news websites we analysed.
Consequently, fake news websites rely on popular and credible ad networks to generate revenue. 
It is interesting to note that before starting such a business relationship between an ad network and a website, there is a vetting process to be followed. 
For example, Google's AdSense ensures that the website complies with its policy~\cite{adsenseSignUp}.
One might expect that during the review process previously described, the popular ad networks would not approve requests of fake news websites, or of websites proven to publish misinformation.

\subsection{Business Relationships}

Although \adstxt files provide a clear view of the ad ecosystem in the analysed fake news websites, this view is based on data provided by the fake news websites themselves.
To provide the point of view of the sellers, we utilize \sellersjson files as provided by the advertising services.
\textbf{\sellersjson} is a complementary mechanism to \adstxt introduced by the IAB Tech Lab to oppose ad fraud and profit from counterfeit inventory.
Specifically, \sellersjson files (format shown in Figure~\ref{fig:sellersJson}) can be used by buyers to discover the final sellers of a bid request (either direct sellers or intermediaries).
In an attempt for a more transparent marketplace, each seller (\ie SSP) publishes in its own \sellersjson file all entities, with which it has business relationships.
According to the specification~\cite{sellersJsonStandard}, the list of entities which are represented by the ad network must be included in this file, even if their identity is confidential.
For that reason, a seller ID is required.
This ID is the same as the one that appears in the website's \adstxt file.
Finally, for each seller ID, the type of the account must be specified
(i) as \texttt{PUBLISHER}, if ad inventory is sold on a website directly owned by the company and the ad network pays the company directly,
(ii) as \texttt{INTERMEDIARY}, if ad inventory is sold by an entity which does not directly own it.
Using this information, we are able to extract reliable information and match it against the one provided by websites to ensure that there are no falsely listed business relationships.

To verify the business relationships we found in \adstxt files, on 12 Jan 2022, we download and parse the \sellersjson files of all popular ad services found in our previous analysis.
We exclude \emph{sovrn.com} from this experiment as we were unable to retrieve its \sellersjson file.
For each identifier with \texttt{DIRECT} relationship found in \adstxt files of news websites, we verify whether the respective business relationship is also registered by the advertising system in its \sellersjson file.
Note that we do not investigate whether there is a relationship mismatch since they are considered as out of scope for this work and left for future research.
Instead, we focus on whether there is a business relationship of any kind between a news website and the respective ad network.

In Figure~\ref{fig:sellersJsonDirect}, we present our findings for the top 10 most popular sellers.
We find that for all ad networks, the results reported in Figure~\ref{fig:adsTxtDirect} and Figure~\ref{fig:sellersJsonDirect} are very similar (or even the same). 
We attribute the small disparities between the two figures to
(i) the fact that \adstxt files might not be all-inclusive, up-to-date or syntactically correct~\cite{bashir2019longitudinal}, and
(ii) the common discrepancies and mislabeled relationships between \adstxt and \sellersjson files~\cite{adsTxtDiscrepancies1}.

Despite the differences that may exist, the important thing to focus is that both points of view agree.
A substantial percentage of fake news websites receive ads through well-known services including 
\emph{google.com}, \emph{indexexchange.com}, \emph{appnexus.com}, \etc
Even though according to Google's Terms of Service, content that makes false claims or contradicts scientific consensus is not eligible for monetization~\cite{googlePublishingPolicy}, this is not the case.
74.3~-~80.8\% of the fake news websites in our analysis have a \texttt{DIRECT} relationship with \emph{google.com} (\ie receive ads through Google), 47.0~-~49.0\% with \emph{indexechange.com}, and 46.0~-~52.5\% with \emph{appenexus.com}.
Please note that compared to previous work~\cite{zeng2020bad}, these findings have not been inferred or detected using a custom methodology.
The importance of these results along with the ones presented in Section~\ref{sec:adsTxt}, is that they are reported by the involved entities themselves.

\point{Finding:} It is evident that news websites tend to form business relationships with ad networks in order to monetize their published content and generate revenue.
Based on our analysis, we find that not all such networks evaluate their clients or refuse deals with fake news websites.
Such ad companies prefer to increase their profits at the expense of a more transparent, reliable and safe Web.
Therefore, even if these business relationships have been formed due to lack of thorough examination of news websites, it is evident that some ad networks facilitate fake news content on the Web.
\section{Advertising on Fake News Websites}
\label{sec:whoFunds}

\subsection{Ad Detection}
\label{sec:adDetection}
\begin{figure}[t]
    \centering
    \includegraphics[width=1\columnwidth]{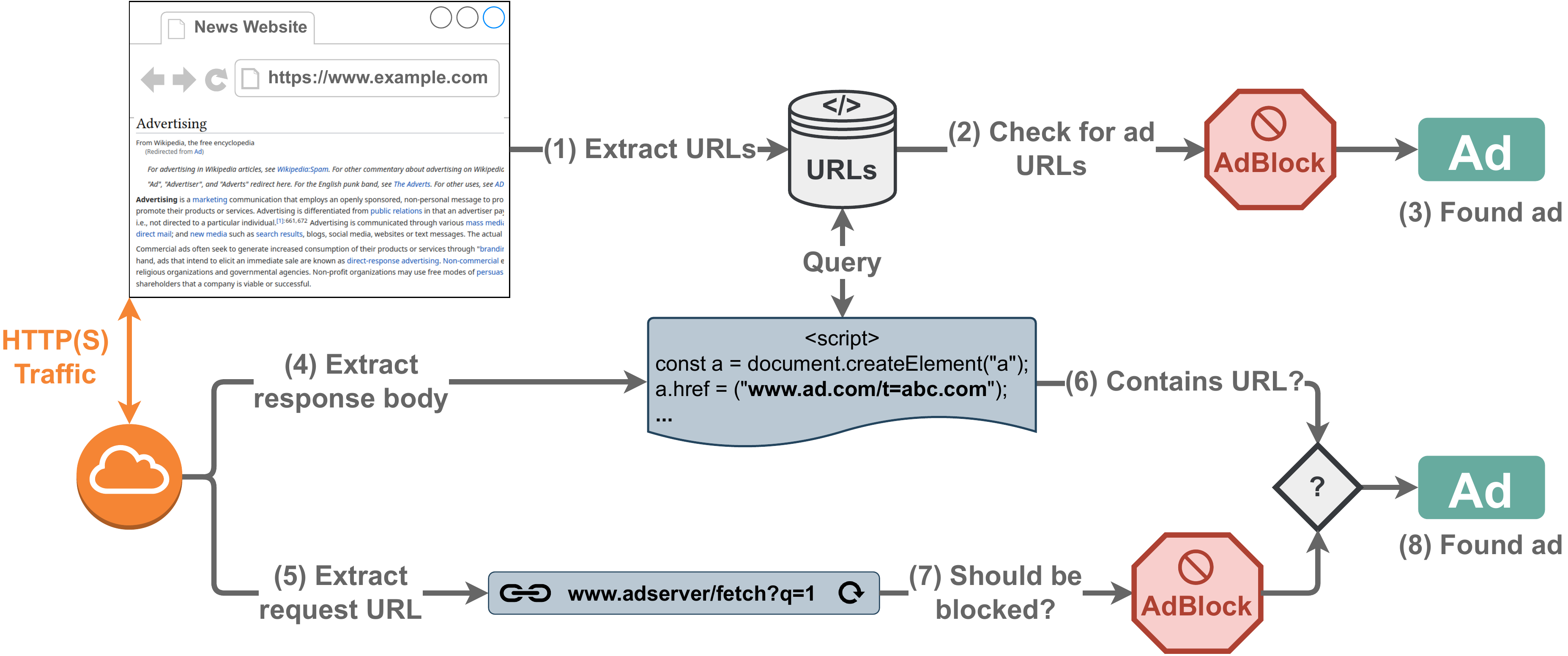}
    \caption{Ad detection methodology that combines both external block lists and network traffic monitoring.}\vspace{-0.4cm}
    \Description{Diagram of the ad detection methodology presented in this work. There are two main components. One that represents external blocking lists and one for network traffic monitoring. In each component there are blocks that indicates the steps followed before deciding if a URL is an ad or not. The steps are (1) Extract URLs, (2) Check for ad URLs, (4) Extract response body, (5) Extract request URL, (6) Contains URL and (7) Should be blocked. The steps (3) and (8) are the outputs of the external block list and network traffic monitoring components that indicate whether a URL is an ad or not.}
    \label{fig:adDetectionMethodology}
\end{figure}

Detecting ads embedded in websites is not trivial~\cite{sjosten2020filter}.
The main difficulty is that the final advertiser may be selected after an auction and is accessed after several re-directions. 
To detect ads embedded in websites and identify the actual advertisers, we propose and implement a novel methodology as outlined in Figure~\ref{fig:adDetectionMethodology}.
The novelty of this methodology lies in the fact that it consists of two distinct components: \textit{external blocking lists} and \textit{network traffic monitoring}.

First (step 1), we extract all URLs from the landing page of the website.
Using the Chrome DevTools protocol, we extract all hyperlinks that can be found even in iFrames, where ads are most commonly found, or the Shadow DOM.
From the extracted URLs, we consider only URLs to other domains.
Next (step 2), we search for URLs belonging to ad networks and represent ads.
When users click on such URLs, either directly, or because they clicked on an image, they are redirected to the advertiser's landing page.
We make use of Brave's adblock engine~\cite{braveAdblock} and the popular open-source filter lists \emph{EasyList}~\cite{easyList} and \emph{uBlock Origin}~\cite{uBlockOrigin} to evaluate URLs, and detect (step 3) those which are ads.

Additionally, our methodology is able to detect ad URLs that belong to the actual advertiser, and not to an ad network that redirects to the advertiser.
We perform an application-level network traffic analysis and trace HTTP(S) requests.
For each request, we extract the body of the response (step 4) and the request URL (step 5).
For a more robust and thorough approach, we follow all redirect chains.
If we find a URL in the response, and we know that this URL has been placed in the website (step 6), we determine whether the original request was towards an advertising domain (step 7), using EasyList and uBlock Origin filter lists.
If so, we deduce that this URL has been placed into the website through an ad network, and consequently, it is an ad URL (step 8).
By combining the two approaches (steps 1, 4, and 5), our methodology is able to detect ad URLs that are either direct URLs to the actual advertiser, or URLs of ad networks that eventually redirect to the advertiser.
To establish and attribute the actual advertiser, we navigate to the detected ad URLs and extract the landing page.

\point{Manual Verification:} To validate our methodology, we use a list of popular websites.
Using \emph{SimilarWeb}~\cite{similarWeb}, we extract the 50 most popular websites from the ``News and Media'' and ``Sports'' categories, for a total of 100 websites.
We select these categories based on empirical analysis, since they are more likely to contain ads.
We apply our methodology for ad detection and advertiser attribution on these websites, while at the same time storing a screenshot of the website. 
Next, we manually evaluate how accurately our method can detect ads on these 100 websites.
We find that our approach has both high Precision ($92\%$ of ``ads'' marked in the websites are actual ads), and Recall ($87\%$ of actual ads in the websites were correctly detected).
These results indicate that our method detects very accurately most ads in websites, with very few false positives.
To ensure the reproducibility of our study, we release a collection of annotated screenshots with ads detected by our methodology~\cite{openSourceData}.

\subsection{Who buys ad space on fake news sites?}
Using our ad detection methodology, we extract the actual entities that advertise in news websites.
Using a clean browser state (\ie no synthetic personas), we visit each ad that our methodology detects and extract the domain of the advertiser.
Our methodology was able to detect $\sim$900 distinct advertisers in real news websites and $\sim$200 advertisers in fake news websites.

We find that a considerable number of fake news websites does not monetize their content via ads.
This is inline with the finding of Section~\ref{sec:adsTxt} and is further discussed in Appendix~\ref{sec:fake-news-ads}.
However, on those who do, we discover that entertainment advertisers with captivating and luring ads are the most popular ads.
In particular, \emph{newscityhub.com} and \emph{inspiredot.net} are the most popular advertisers, appearing in 15\% and 14\% of fake news websites, respectively.
These advertisers are known for using click-bait ads with ``catchy'' titles that entice the visitor's curiosity. 
Even unintentionally, these advertisers are common among the misinformation websites we study and provide great revenue to their operators.
Please note that often, advertisers have control over where their ads will appear and, therefore, share a portion of the ethical responsibility for the proliferation of fake news content.
For example, in the Google Ads platform, advertisers can choose where their ads are displayed~\cite{googleAdsAppear} and even exclude specific websites~\cite{googleAdsExclude}.
Similarly, Rubiconproject (now called Magnite) respects advertisers' blocklists regarding where their ads will appear~\cite{magniteBlocklists}.

Next, we extract the categories of advertisers by utilizing \emph{Cyren}~\cite{cyren}, whose classification engine has already been used in previous academic works (\eg~\cite{papadopoulos2017long,carrascosa2015always,diamantaris2020seven}), and has been proven that it can classify a greater set of websites than other similar systems~\cite{carrascosa2014understanding}.
Using their classification service, we are able to extract the categories of over 95\% of advertisers in our dataset.
For websites assigned to multiple categories, we single out the most frequent label in our dataset.
Figure~\ref{fig:advertisersCategories} illustrates the types of distinct advertisers collaborating with real news and fake news websites.

The majority of advertisers in both fake and real news websites come from the ``Business'' category.
This behavior is expected, since these advertisers promote websites that contain business-related information in an attempt to popularize their services or products.
Also, we observe that a large number of fake news websites (almost 40\%) display ads from the ``Entertainment'' websites.
These ads contain captivating, and, sometimes even click-bait, content from celebrity websites, television and movie programs, as well as entertainment news that tempt users.
The rest of the advertisers fall into ``Technology'', ``Shopping'', ``Education'', ``News'', \etc
On the other hand, real news websites place ads coming primarily from advertisers of other businesses, news, and education-related services.
``Spam'' category seems less prominent in fake than real news sites.

\point{Finding:}
We observe that click-bait and captivating ads are more likely to appear on fake news websites.
Such advertisers fuel fake news content and, through their ad impressions, financially support part of the ecosystem.
We also find that advertisers on fake news sites seem to be normal and legitimate business.
Our results suggest that fake news websites host ads from legitimate advertisers, thus doing serious ad business and avoiding ads from malicious or dodgy sites such as SPAM, which could risk their monetization avenues or jeopardize their existence in the ad ecosystem.

\section{News Websites Ownership}
\label{sec:clusters-analysis}

In this section, our goal is to answer: \emph{Who owns fake news websites?} and \emph{What other websites do the owners of fake news websites operate?}
Towards this goal, we expand our dataset, since so far we focused on websites which were clearly categorized as either fake or real news.
Thus, in this analysis, we include a corpus of 1,548 extra news websites from the sources of Section~\ref{sec:lists}, which were not clearly categorized as either fake or real, for a total of 3,960 news websites.

\subsection{Community Detection}
To be able to answer what kind of other websites the owners of fake news websites own, we first need to determine who the owner of a fake news website is.
Although this question is rather tricky to answer, we capitalize on the methodology described in~\cite{coOwnershipGraphs}.
The methodology makes use of four different types of \identifiers used in three separate Google Services.
Contrary to Sections~\ref{sec:facilitators} and~\ref{sec:whoFunds}, the analysis of this section is bound to websites that make use of such Google services.
Then, websites can be linked together if they contain common such identifiers.

\point{Publisher-specific ID detection: }
Such identifiers are alphanumeric values that follow strict formats and uniquely identify user accounts in popular services, such as AdSense and Google Analytics.
Administrators embed these identifiers in their websites in order to use the respective service.
For example, admins need to embed an identifier in the form of \texttt{UA-123456-7} in order to use Google Analytics.
Since some of these identifiers are associated with the receipt of the funds generated via ads, it is generally safe to assume that websites that share the same identifier (\ie give their ad revenue to the same entity) are closely related, or even owned by the same entity~\cite{coOwnershipGraphs}.
Using regular expressions and common data cleaning techniques, identifiers are extracted from the HTML code of websites, network traffic and first- or third-party cookies.
Then, values that are words of the English dictionary, or match a custom list of common keywords are removed.
Table~\ref{tab:dataset} summarizes the websites containing \identifiers.
We find that there are 385 fake news websites and 1,025 real news websites with at least one type of identifier.
A rundown of the detected identifiers can be found in Appendix~\ref{sec:cluster-composition}.
We find that for most types of identifiers, there are more domains than actual identifiers, indicating that there are identifiers being re-used in more than one domain. 

\begin{table}[t] 
    \centering
    \scriptsize
    \begin{tabular}{lrrrr}
    \toprule
        \textbf{Description}    & \textbf{Volume}   & \textbf{\% of total} & \textbf{FN} & \textbf{RN}\\ 
    \midrule
        Initial set of websites                 & 3,960   & 100.00\% & - & -\\
        Websites successfully crawled           & 3,311   & 83.61\% & - & -\\
        Websites that errored                   &   649   & 16.39\% & - & -\\
    \midrule
        Websites with no ad-related identifiers &   737   & 22.26\% & 325 & 172\\
        Websites with at least one identifier   & 2,574   & 77.74\% & 385 & 1,025\\
        Websites with all types of identifiers  &   184   &  5.56\%  & 2 & 62\\
    \bottomrule
    \end{tabular}
    \caption{Summary of crawled News websites. ``FN'' stands for fake news websites while ``RN'' stands for real news websites.}
    \vspace{-0.6cm}
    \label{tab:dataset}
\end{table}

\begin{figure*}[t]
    \centering
    \hfill
    \begin{minipage}[t]{0.323\textwidth}
        \centering
        \includegraphics[width=1\columnwidth]{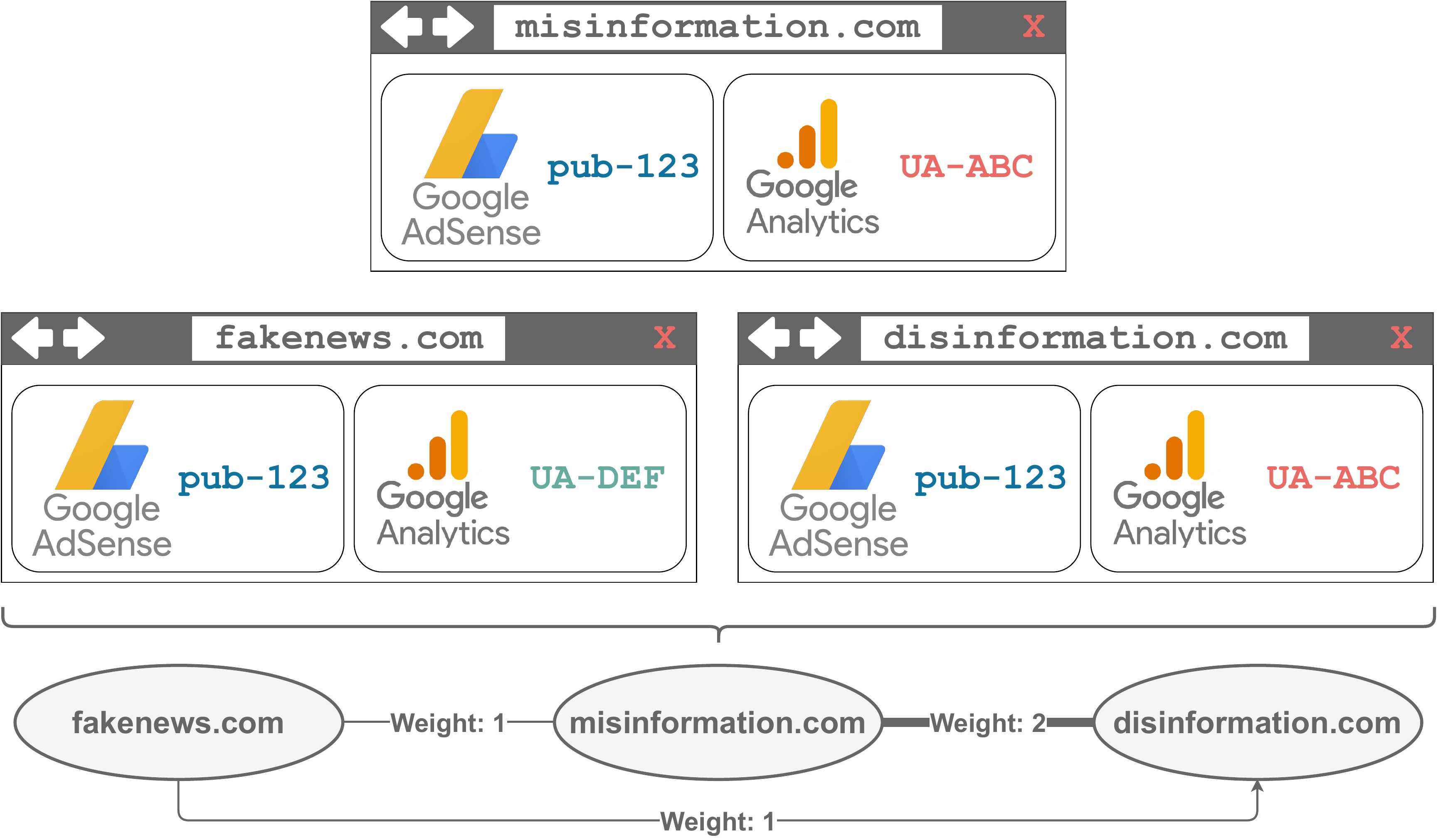}
        \caption{Example of a Metagraph construction. Websites that share identifiers are linked together in the resulting graph. The more identifiers they share, the greater the edge weight.}
        \vspace{-0.4cm}
        \Description{Diagram that illustrates how publisher IDs of different websites are combined to form a Metagraph. There are three different websites. misinformation.com contains the identifiers "pub-123" and "UA-ABC", fakenews.com contains the identifiers "pub-123" and "UA-DEF" and disinformation.com contains the identifiers "pub-123" and "UA-ABC". This information leads to a Metagraph with three nodes, one for each website. There is an edge between disinformation.com and misinformation.com, with weight 2 while fakenews.com is linked to misinformation.com and disinformation.com with edges of weight 1.}
        \label{fig:graphMethodology}
    \end{minipage}
    \hfill
    \begin{minipage}[t]{0.323\textwidth}
        \centering
        \includegraphics[width=1\columnwidth]{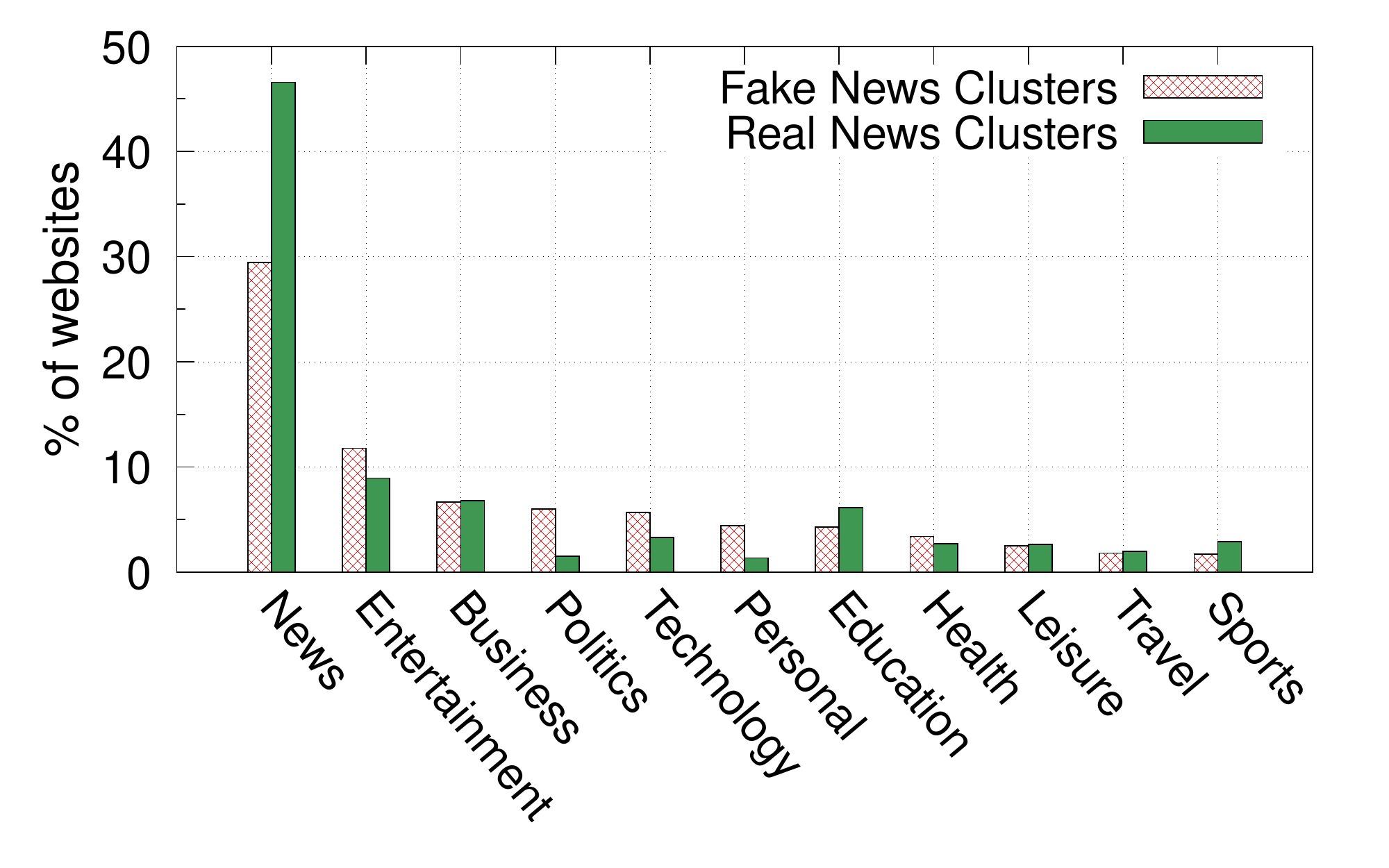}
        \caption{Distribution of website categories in each cluster. We see that most sites in a cluster are ``News'', followed by ``Entertainment'', and ``Business''.}
        \Description{Histogram illustrating the categories of websites found in communities that contain fake or real news websites. Each bar represents the percentage of websites of the respective category. The categories are News, Entertainment, Business, Politics, Personal, Education, Health, Leisure, Travel and Sports.}
        \label{fig:websiteCategories}
    \end{minipage}
    \hfill
    \begin{minipage}[t]{0.323\textwidth}
        \centering
        \includegraphics[width=1\columnwidth]{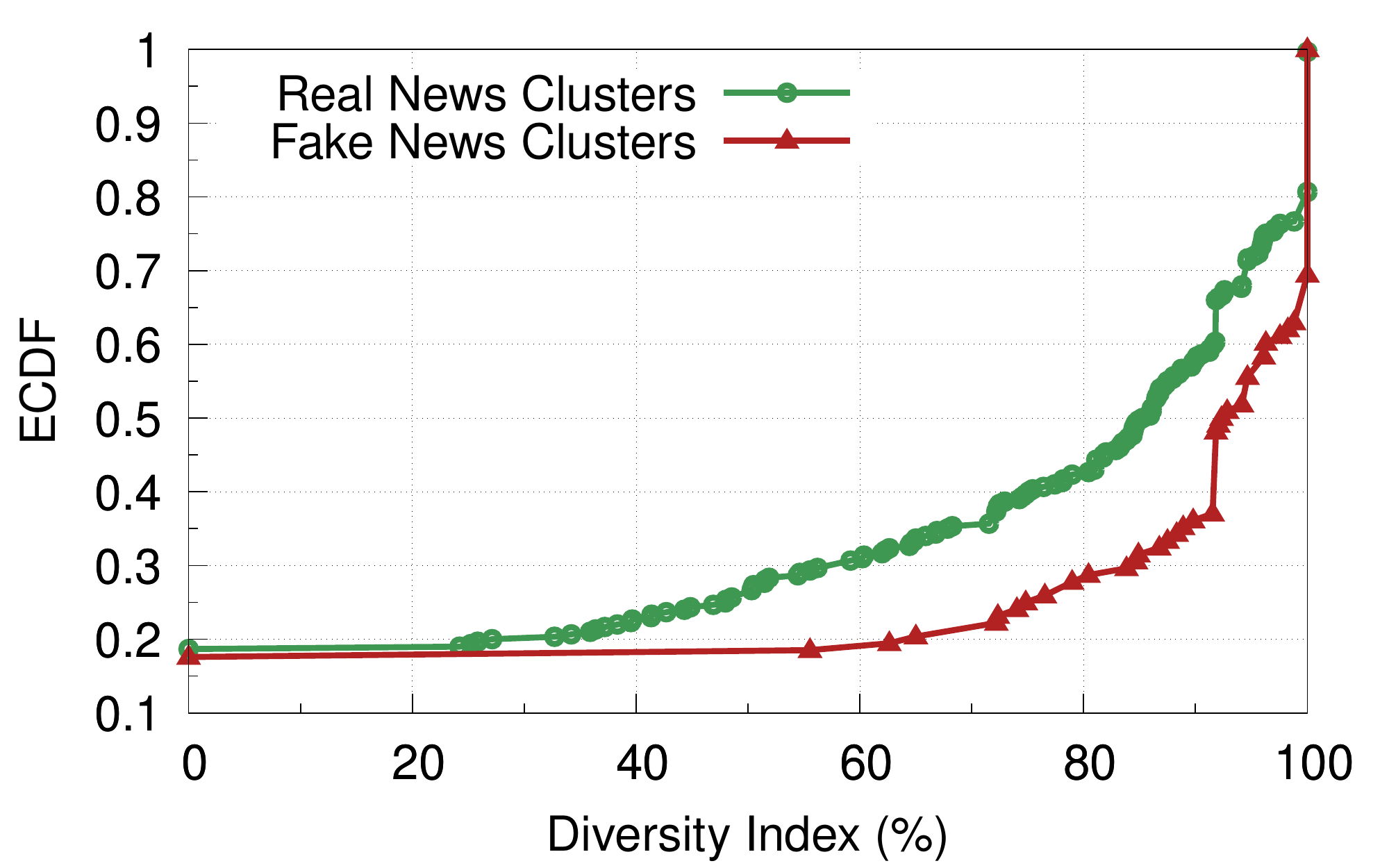}
        \caption{Diversity score based on Shannon's Diversity Index. Real News (green circles) seem to cluster higher from Fake News (red triangles) indicating more homogeneity.}
        \Description{Cumulative Distribution Function of Shannon's Diversity Index for the communities that contain fake or real news websites. There are two different lines, one for fake news clusters and one for real news clusters. The line for real news clusters is higher than the one for fake news clusters.}
        \label{fig:diversityIndex}
    \end{minipage}\vspace{-0.3cm}
\end{figure*}

\point{Graph Analysis \& Cluster Construction: } Using the aforementioned detected identifiers, we construct a~\emph{Metagraph}, a graph that represents the relationships among websites.
This graph contains only website nodes and those that share an identifier (that uniquely identifies an account) are connected through an edge.
The weight of the edge is proportional to the number of identifiers websites share.
A large edge weight represents greater confidence that these two websites are indeed operated and managed by the same entity.
The notion of the Metagraph has been assessed in~\cite{coOwnershipGraphs}, where the authors show that it can accurately detect websites operated by the same entity and validated its performance against other  techniques.
Figure~\ref{fig:graphMethodology} illustrates the construction of such a toy Metagraph.

To detect clusters of websites operated or even owned by the same entity, a graph community detection algorithm is applied on the Metagraph.
Contrary to~\cite{coOwnershipGraphs} which uses the Girvan-Newman method, in this work we apply the Louvain method~\cite{blondel2008fast}.
Our decision is only based on the performance benefits of the Louvain method: it is faster, scalable, and able to accommodate the entire Metagraph without performing any edge-pruning.
Additionally, we integrate information from the \trancoCrawl dataset~\cite{coOwnershipGraphs}, which contains \identifiers found in the top 1M most popular websites of April, 2021.
The resulting Metagraph contains over 114.5K website nodes and 443K edges.
More information about the community detection process can be found in Appendix~\ref{sec:cluster-composition}.

We define a \emph{Fake news cluster} as a community of websites that contains at least one fake news website.
This implies that a community is operated or owned by an entity which, among other business, also spreads fake news.
Similarly, we define a \emph{Real news cluster} as a community with at least one real news website.
It is worth noting that, in these definitions, we do not label other websites inside the clusters, we simply characterize the clusters they belong to.

\subsection{Categories of website clusters}
\label{sec:cluster-categories}

By definition, each Fake news cluster contains at least one fake news website.
However, it also contains other websites as well. 
Figure~\ref{fig:websiteCategories} shows the types of other websites contained in each such cluster.
We see that for the Fake news clusters (red bars) about 29.5\% of the websites are news.
The rest (almost 70\%) are ``not news''  websites and encompass ``Entertainment'', ``Business'', ``Politics'', ``Technology'', \etc
This verifies that most fake news website owners are also engaged in other types of businesses.
Contrary, we observe that entities that own or operate real news websites, also tend to manage other news websites in order to reach a wider audience or even convey other types of news.
It seems that both types of cluster have some diversity, but it is not clear which is more diverse. 

To clarify this, we study Shannon's diversity index~\cite{nolan2006beachcomber}, a statistical measure that can indicate how many different categories there are in a community, while at the same time reflecting the relative abundance of website categories.
Shannon's diversity index is defined as ${H}' = -\sum_{i=1}^{S} p_i \ln p_i$, where $S$ is the number of different categories in the dataset (\ie richness) and $p_i$ is the proportion of websites belonging to category $i$.
When all categories in a community are equally common, the Shannon index takes the maximum value $ln(R)$.
The more unequal the categories are, the smaller the index is.
Shannon's diversity index equals zero when there is only one category of websites in a community.

We apply this statistical measure to communities that contain fake news or real news websites.
Figure~\ref{fig:diversityIndex} illustrates the distribution of the diversity index for fake news and real news clusters.
This index is normalized by $ln(R)$, which is the case where all categories are equally common.
Consequently, the case of 0\% in Figure~\ref{fig:diversityIndex} indicates that there is only one category of websites in the community, while the case of 100\% suggests that the categories are equally distributed, thus revealing a diverse community.
We see that Real news clusters tend to cluster higher and to the left (for the same value of $y$) of the Fake news clusters.

\point{Finding:} We find that Real news clusters are more homogeneous: owners of these clusters tend to focus on a smaller number of different Web businesses.
At the same time, owners of Fake news clusters seem to engage in higher diversity in their business. 
Combined with the fact that fake news website owners have a preference towards ``Entertainment'' and ``Business'' websites, we speculate that their goal is to monetize their websites and generate revenue, and that fake news websites might be a way to make ``quick buck''.

\subsection{Who owns fake news websites?}
\label{sec:cluster-ownership}

In order to study fake news websites owners, we manually investigate communities\footnote{For the communities, we rely on the accuracy of the methodology as presented in~\cite{coOwnershipGraphs}.} that contain at least one fake news website, and discover the legal entity that operates the websites of each community.
To provide a better understanding of the fake news ecosystem and highlight its social effect, we selectively report some of these communities.
More information about the methodology we follow along with other striking examples of fake news websites ownership can be found in Appendix~\ref{sec:ownership-examples}.
We make the clusters of fake news websites publicly available~\cite{openSourceData}.

We detect a community of websites related to the \emph{Family Research Council} (FRC), an activist group with an affiliated lobbying organization.
However, one of these websites has been labeled as a ``Questionable Source'' by MBFC since it promotes far-right propaganda, it lacks transparency regarding funding and it has numerous failed fact checks~\cite{frcMBFC}.
The Southern Poverty Law Center (SPLC) designated FRC as a hate group~\cite{hateGroup}.
Additionally, we discover the websites \emph{thetruthaboutcancer.com} and \emph{thetruthaboutvaccines.com}, owned by Ty and Charlene Bollinger.
Their websites promote both unproven and dangerous remedies (\ie pseudoscience), as well as information regarding \covid and vaccines which has been proven to be false~\cite{bollingerMBFC}.
In fact, Ty and Charlene Bollinger have been identified as part of the ``Disinformation Dozen'', a set of 12 individuals that produce 65\% of the misinformation and misleading claims regarding \covid on social media~\cite{nprDozen, disinformationDozen}.

\point{Finding:} These examples, along with others excluded for brevity, demonstrate the correctness and efficiency of our methodology.
That is, we are able to accurately detect communities of fake news websites, owned or operated by the same entity that pushes a specific political or ideological agenda, and tries to shift the public opinion.
In fact, people may be led to accept false beliefs or even make life-altering decisions based on this false information~\cite{reutersFakeNews}.
We believe our methodology can play a vital role in this problem: if a person can be informed that the website they are visiting is owned by an entity that also operates or owns fake news site(s), the visitor will most likely view the content with more caution.
\section{Discussion \& Conclusion}
\label{sec:conclusion}

\subsection{Limitations}
Even though we study what the Web ad-ecosystem uses in its majority~\cite{bashir2019longitudinal}, we understand that there are limitations to \adstxt files~\cite{adsTxtDiscrepancies1, adsTxtDiscrepancies2}.
Moreover, the analysis of advertisers relies on the methodology presented in Section~\ref{sec:adDetection}, and, though our methodology has a high precision score, we acknowledge that it might fail to detect some ads.
Also, our network monitoring approach will miss fragmented ad URLs.
These limitations do not reduce the credibility of our findings, since we  still study a big portion of the ad ecosystem.
Furthermore, we made efforts to exclude intermediary publishing partners from communities of websites operated by the same entity, as discussed in~\cite{coOwnershipGraphs}.
Finally, domain classification services might suffer from classification and disagreement flaws and no service is error-free~\cite{vallina2020mis}.
We choose Cyren because it
(i) accepts miss-classification reports,
(ii) is language and content agnostic, and
(iii) has a vast database of 140 million classified domains.

\subsection{Summary}
The success of curbing fake news primarily depends on the ability of stakeholders to remove the incentives of fake news producers.
One may think that fake news sources are supported only by shady organizations enlisting people in remote countries~\cite{teensInBalkans} and legitimate ad-networks have pulled out from such misinformation sources.
In this work, we show that this, unfortunately, is not the case.

We identify and study the companies that advertise in fake news websites and the middlemen responsible for keeping the avenues of ad revenue open.
We show that popular, legitimate advertising systems (such as Google, Indexexchange and AppNexus) have a \emph{direct} advertising relation with more than 40\% of the fake news websites in our list.
Through clustering based on advertiser IDs present in such websites, we report that operators of fake news sites usually operate a set of websites that include entertainment, business, politics, \etc
This indicates that the operation of a fake news website is part of a larger business and not an isolated event.

We believe that the Metagraph described in~\cite{coOwnershipGraphs} and used in this work provides clear understanding of relationships among websites.
We plan on exploiting the sensitive information related to advertising and analytics services to develop a content-agnostic classifier that can automatically detect fake news websites.
Contrary to common content-aware fake news detection schemes and manual fact-checking campaigns, we believe that such a classifier can effectively detect fake news websites that have just spawned through the entities that own or operate them.
\begin{acks}
This project received funding from the EU H2020 Research and Innovation programme under grant agreements No 830927 (Concordia), No 830929 (CyberSec4Europe), No 871370 (Pimcity), No 871793 (Accordion), No 101021808 (Spatial), and No 883543 (CC-DRIVER).
These  results reflect only the authors' view and the Commission is not responsible for any use that may be made of the information it contains.
\end{acks}

\bibliographystyle{ACM-Reference-Format}
\balance
\bibliography{main}

\appendix
\section{Ethical Considerations}
\label{sec:ethics}

The execution of this work has followed the principles and guidelines of how to perform ethical information research and use of shared measurement data~\cite{dittrich2012menloreport,rivers2014ethicalresearchstandards}.
We keep our crawling to a minimum to ensure that we do not slow down or deteriorate the performance of any web service in any way, and make concerted effort not to perform any type of DoS attack to the visited website.
Therefore, we crawl only the landing page of each website and visit it only once.
We do not interact with any component inside a website, and only passively observe network traffic.
Consequently, we emulate the behavior of a normal user that stumbled upon a website.

In accordance to the GDPR and ePrivacy regulations, we did not engage in collection of data from real users.
Also, we do not share with any other entity any data collected by our crawler.
We intentionally do not make our crawled dataset public (but only the fake and real news lists), to ensure that there is no infringement of copyrighted material from any website.

Finally, regarding the ad detection methodology, we were cautious not to affect the advertising ecosystem or deplete advertiser budgets.
The development and testing of our methodology was performed on offline captures of websites.
Additionally, for each website we process, we visit only the landing page and ``click'' on advertisements only once.
\section{Ad Networks}
\label{sec:adNetworks}

Complementary to the analysis of Section~\ref{sec:adsTxt}, we examine how many fake news websites have a \texttt{RESELLER} relationship with the ad networks studied so far. 
A \texttt{RESELLER} business relationship expresses cases where a third party has been authorized to control the ad space~\cite{adsTxtStandard}.
Table~\ref{tab:adsTxtReseller} presents the results. 
We find that 67.71~-~73.73\% of fake news websites in our dataset have a \texttt{RESELLER} relationship with \emph{appnexus.com}, \emph{openx.com}, \emph{rubiconproject.com}, \emph{indexechange.com}, and \emph{pubmatic.com}.
We note that these percentages reported in Table~\ref{tab:adsTxtReseller}) are even higher than those reported in Figure~\ref{fig:adsTxtDirect}.
For example, although as many as 52.5\% of fake news websites engage in a \texttt{DIRECT} relationship with \emph{appnexus.com}, an even higher percentage of them (73.73\%) engage in a \texttt{RESELLER} relationship with it.
The same trend is true for the rest of the ad networks, which means that roughly six out of ten fake news websites have \texttt{RESELLER} relationships with the major ad networks. 

\begin{table}[h] 
    \centering
    \scriptsize
    \begin{tabular}{lr|lr}
    \toprule
        \multicolumn{2}{c|}{\textbf{Real News}} & \multicolumn{2}{c}{\textbf{Fake News}} \\
        \textbf{Service} & \textbf{Portion} & \textbf{Service} & \textbf{Portion} \\ 
    \midrule
        appnexus.com       & 86.92\% & appnexus.com       & 73.73\% \\
        openx.com          & 85.32\% & rubiconproject.com & 69.19\% \\
        rubiconproject.com & 85.00\% & pubmatic.com       & 68.68\% \\
        indexexchange.com  & 85.00\% & spotxchange.com    & 67.67\% \\
        pubmatic.com       & 84.37\% & spotx.tv           & 67.17\% \\
    \bottomrule
    \end{tabular}
    \caption{Most popular ad networks with \texttt{RESELLER} relationships. Publishers authorize intermediary entities to operate their accounts.}
    \label{tab:adsTxtReseller}
\end{table}
\section{Cluster Composition}
\label{sec:cluster-composition}

For the construction of the Metagraph we make use of the detected \identifiers (Table~\ref{tab:identifiers}).
However, very large communities of websites may arise due to the presence of intermediary publishing partners.
These are third-party services that help publishers manage their websites and increase website popularity, and consequently generate more revenue.
In this work, we focus only on identifiers which can be found in more than 1 but at most 50 websites.
We use this threshold based on the analysis of~\cite{coOwnershipGraphs}, declaring these as \emph{Small} and \emph{Medium} classes of website administrators.
Larger clusters are considered intermediary publishing partners and not an actual administrator or an owner~\cite{googlePartners}.
The use of these two classes in the Metagraph eliminates the issue of intermediary partners.

\begin{table}[h] 
    \centering
    \scriptsize
    \begin{tabular}{lrrr}
    \toprule
        \textbf{Description}    & \textbf{Unique}   & \textbf{Unique Domains} & \textbf{\% successful} \\
        & \textbf{Identifiers} & \textbf{of landing URLs} & \textbf{websites} \\
    \midrule
        \publisherIDs   &   642 &   872 & 26.34 \\
        \trackingIDs    & 2,638 & 2,365 & 71.43 \\
        \measurementIDs &   393 &   584 & 17.64 \\
        \containerIDs   & 1,113 & 1,221 & 36.88 \\
    \bottomrule
    \end{tabular}
        \caption{Detected \identifiers in news websites.}\vspace{-0.3cm}
    \label{tab:identifiers}
\end{table}

In order to detect communities of websites operated by the same entity, we employ the Louvain community detection algorithm~\cite{blondel2008fast}.
We perform hierarchical clustering by successive instances of the algorithm, and extract a dendogram, where each level is a partition of the metagraph nodes.
Level 0 contains the smallest communities while moving to higher levels results to bigger communities.

Table~\ref{tab:communities} summarizes the detected communities of websites operated by the same entity.
We observe that for higher levels of the dendogram (\ie levels 1 and 2), fake news clusters contain thousands websites, and each such cluster is very big in size (\ie  50.43 for level 2).
Communities in higher levels are formed due to the presence of intermediary publishing partners that control hundreds or even thousands of websites and according to~\cite{coOwnershipGraphs}, do not indicate a clear co-administration relationship.
Thus, we focus only on the first level of the dendogram, containing small and more accurate communities.
We find 73 fake news clusters that remain identical across different dendogram levels.

\begin{figure}[h]
    \centering
    \includegraphics[width=.9\columnwidth]{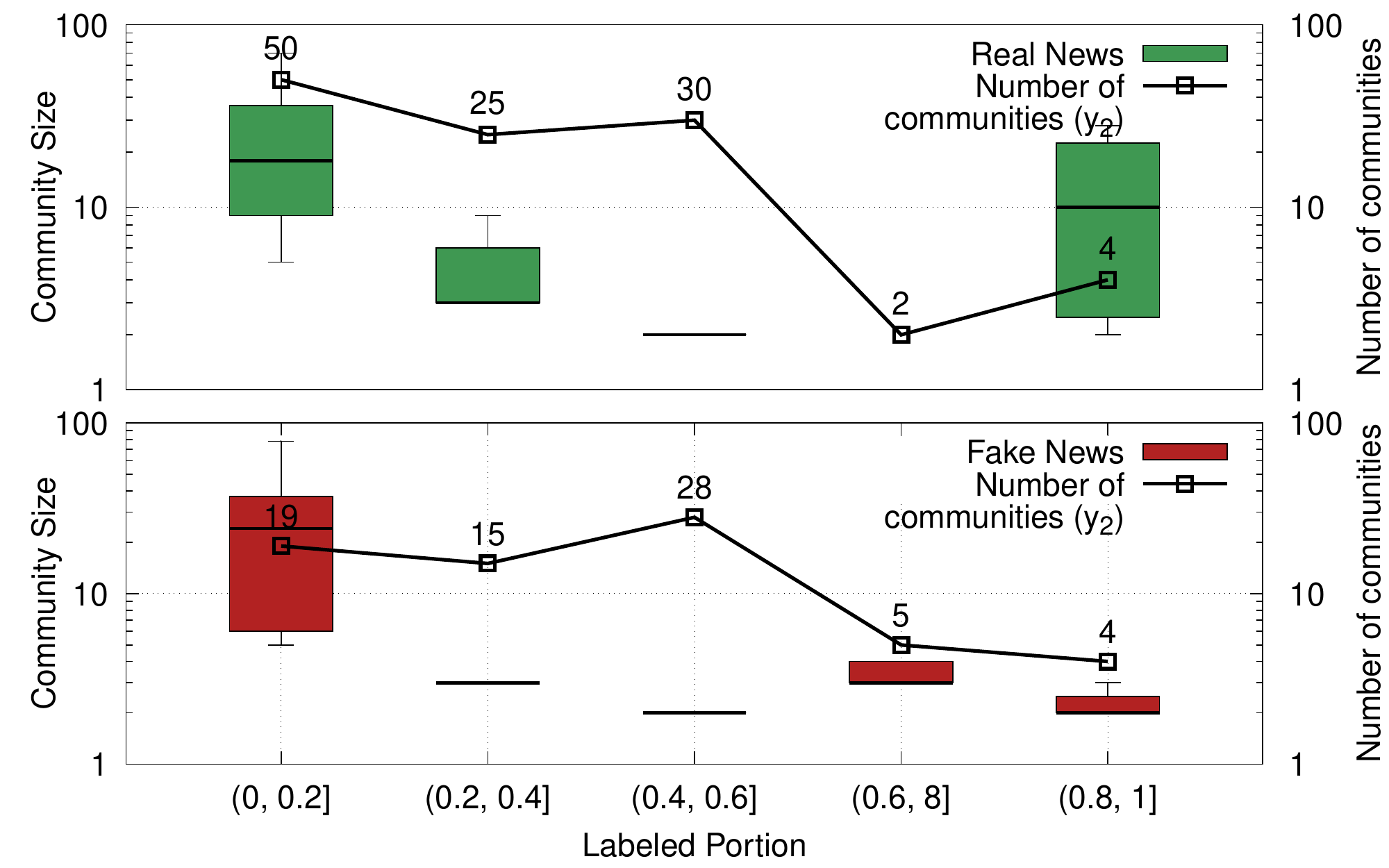}
    \caption{Distribution of sizes of detected clusters and portion of websites labeled as fake or real news. Real news sites cluster more together (green rectangle, range (0.8, 1]).}
    \label{fig:clustersSizePortion}
\end{figure}

For each identified cluster (Fake news and Real news) we compute the portion of its websites labeled (fake or real, respectively) based on our lists.
For example, a portion of $0.5$ for a Fake news cluster indicates that half of the websites in the cluster were labeled as fake news.
In Figure~\ref{fig:clustersSizePortion}, we illustrate the size of the detected fake news and real news clusters, as well as their labeled portions.
We see that both Fake and Real news clusters behave similarly apart from the (0.8, 1] range.
Indeed, in that range, large communities of real news (the big green rectangle at right) show that we have Real news clusters of decent size where more than 80\% of them are categorized as real news.
This implies that such clusters contain several websites that disseminate credible information.
On the contrary, for the Fake news clusters, in the same range (\ie (0.8, 1]) we see that the red rectangle is very thin with a value close to two.
This implies that fake news websites do not tend to cluster together - at least not as much as the real news websites do. 
These results are inline with the ones presented in Section~\ref{sec:cluster-categories}.

\begin{table}[] 
    \scriptsize
    \centering
    \vspace{-0.2cm}
    \begin{tabular}{lrrrr}
    \toprule
        \textbf{Level} & \textbf{Number of} & \textbf{Fake News} & \textbf{Websites in} & \textbf{Average}\\
        & \textbf{communities} & \textbf{clusters} & \textbf{clusters} & \textbf{Cluster size}\\
    \midrule
    0 & 32,961 & 108 &   883 &  8.17 \\
    1 & 30,832 &  95 & 2,843 & 29.92 \\
    2 & 30,687 &  91 & 4,590 & 50.43 \\
    \bottomrule
    \end{tabular}
    \caption{Detected website communities using Louvain method.}
    \vspace{-0.4cm}
    \label{tab:communities}
\end{table}
\section{Fake news websites owners}
\label{sec:ownership-examples}

In this section, we provide some examples of fake news sites, their owners, and their ecosystem.
We establish the entity by manually reviewing the copyrights claim, the privacy notice or the terms of services provided voluntarily by the websites themselves.
We do not utilize external resources to ensure that any information about the ownership of a website is willingly provided by their administrators.
We focus on smaller clusters, for which we have greater confidence about their miss-informative nature, as discussed in Appendix~\ref{sec:cluster-composition}.

For example, we find a cluster of 6 websites published by \emph{Sophia Media}.
4 of these websites are part of the \emph{Health Impact News Network}.
These websites have been marked as ``Pseudoscience websites'' by MBFC, since they promote anti-vaccination propaganda and have multiple failed fact checks~\cite{vaccineImpactMBFC,healthImpactNewsMBFC,medicalKidnapMBFC}.
In fact, in 2020 \emph{NewsGuard}~\cite{newsguard}, a journalism company that tracks online misinformation, identified Health Impact News as one of the greatest spreaders of \covid misinformation on Facebook~\cite{newsGuardSuperSpreaders}.

We also find a set of two websites owned and published by the National Vaccine Information Center (NVIC), an American organization for information about diseases and vaccines.
One of these two websites has been marked as ``Pseudoscience website'' by MBFC since it promotes anti-vaccination propaganda and has multiple failed fact checks~\cite{nvicMBFC}.
Health Impact News was also identified as one of the greatests spreaders of \covid misinformation by Newsguard~\cite{newsGuardSuperSpreaders}.
Moreover, we find the websites \emph{adfmedia.org} and \emph{adflegal.org} being controlled by the same entity with the latter having been labeled as an extreme biased website due to propaganda~\cite{adfMBFC}.
The Alliance Defending Freedom (ADF) is a multi-million organization, which has also been classified as a hate group by the Southern Poverty Law Center (SPLC)~\cite{adfSPLC1}.

Furthermore, we find that not only coordinated organizations, but also individuals are behind communities of fake news websites.
Specifically, we find a pair of websites, \emph{freedomforceinternational.org} and \emph{needtoknow.news}, founded and powered by G. Edward Griffin.
In his websites, he generally promotes right-wing beliefs, but also conspiracy theories and pseudoscience treatments~\cite{griffinMBFC}.
Some of his beliefs about cancer treatment have been debunked by the American Journal of Public Health, since he promoted a banned chemical compound without any scientific evidence~\cite{wwcLandau}.

\point{Ambiguous Website Ownership:} Finally, we observe some contradicting communities that contain both a real (\ie credible) and a fake news website.
These communities are not formed because of a clustering mistake, contrariwise, we show that specific entities operates both types of websites.
First, we observe a community of three websites consisting of \emph{checkyourfact.com}, \emph{smokeroom.com} and \emph{dailycaller.com}.
CheckYourFact, an accepted signatory of the International Fact Checking Network~\cite{checkyourfact}, is considered by MBFC a credible fact checker with high factual reporting that utilizes proper sources and adheres to credible fact checking principles~\cite{checkyourfactMBFC}.
However, according to their \textit{About} page, CheckYourFact is a news product of TheDailyCaller, a conservative news website that deliberately publishes misleading information and false stories~\cite{dailycallerMBFC}.

In addition to this, we also find another contradicting community of 51 different websites.
We manually visited and explored all of these websites and deducted that they belong to \emph{Salem Media Group} and its subsidiaries.
One of the websites in this community, \emph{srnnews.com}, is part of our real news dataset since it has been rated \texttt{HIGH} for its factual reporting and has a clean fact check record~\cite{srnnewsMBFC}.
In the same community, we also find \emph{pjmedia.com}.
This website is labeled as a questionable source since it displays extreme right-wing bias, it regularly promotes propaganda, as well as conspiracy theories, and it has published multiple false stories that failed fact checks~\cite{pjmediaMBFC}.

\section{Ads in Fake News Websites}
\label{sec:fake-news-ads}

In Sections \ref{sec:facilitators} and \ref{sec:whoFunds} we surprisingly discovered that only a portion of fake news websites display digital ads.
This behavior was verified by both the \adstxt files served by the websites themselves, as well as by our ad detection methodology.
To further understand this issue we manually investigated a random subset of 100 fake news websites in our list.
We found that a big number of fake news websites do not display ads because they received funding from various other sources.
For example, both \emph{infowars.com} and \emph{brighteon.com} have online stores, \emph{21stcenturywire.com} and \emph{navarroreport.com} receive funding from publishing magazines and books respectively, while \emph{cosmicintelligenceagency.com} provides paid webinars.
We also discovered a great number of websites that sustain themselves by receiving money from their visitors either through donations (\eg \emph{canadafreepress.com} and \emph{infiniteunknown.net}) or through paid memberships (\eg \emph{aapsonline.org}).
Undoubtedly, there are also these fake news publishers that do not really care about monetizing their content but focus only on pushing their political or ideological agendas (\eg \emph{911truth.org} and \emph{channel18news.com}).
The sources of external funding, which fake news websites receive, are considered a different topic and left for future research.

From a technical point of view, we discovered that some websites do not display ads in their landing pages and only do so if you click on specific articles (\eg \emph{12minutos.com} and \emph{24aktuelles.com}), while others display ads that come from static campaigns that stem from direct business contracts (\eg \emph{abovetopsecret.com}).
Our methodology was not able to handle such cases.
Finally, we found that a great portion of the evaluated websites is no longer active (\eg \emph{24wpn.com} and \emph{embols.com}) and that some websites are no longer maintained and even though they contain ad scripts, these scripts no longer work and cannot fetch ads (\eg \emph{dcgazette.com}).

\end{document}